\documentclass[12pt]{article}
\usepackage{jheppub}

\pdfoutput=1

\usepackage[table]{xcolor}
\usepackage{amsmath,array,amsfonts,graphicx,wrapfig,lscape,float,mathtools,multirow,longtable}
\usepackage{physics}
\usepackage{stackrel}
\usepackage[all]{xy}
\usepackage{subcaption}
\usepackage{url}

\newcommand{\be}{\begin{equation}}
\newcommand{\ee}{\end{equation}}
\newcommand{\beq}{\begin{equation}}
\newcommand{\beql}[1]{\begin{equation}\label{#1}}
\newcommand{\eeq}{\end{equation}}
\newcommand{\ba}{\begin{array}}
\newcommand{\ea}{\end{array}}
\newcommand{\bea}{\begin{eqnarray}}
\newcommand{\beal}[1]{\begin{eqnarray}\label{#1}}
\newcommand{\eea}{\end{eqnarray}}
\newcommand{\ben}{\begin{enumerate}}
\newcommand{\een}{\end{enumerate}}
\newcommand{\bean}{\begin{eqnarray*}}
\newcommand{\eean}{\end{eqnarray*}}
\newcommand{\eref}[1]{(\ref{#1})}
\newcommand{\sref}[1]{\S\ref{#1}}

\newcommand{\fref}[1]{Figure \ref{#1}}
\newcommand{\btab}[1]{\begin{tabular}{#1}}
\newcommand{\etab}{\end{tabular}}

\def \Black {\pdfsetcolor{0 0 0 1}}

\newcommand{\comment}[1]{}

\newcommand{\CN}{{\cal N}}

\newcommand{\qed}{\nobreak \ifvmode \relax \else
      \ifdim\lastskip<1.5em \hskip-\lastskip
      \hskip1.5em plus0em minus0.5em \fi \nobreak
      \vrule height0.75em width0.5em depth0.25em\fi}

\definecolor{darkspringgreen}{rgb}{0.09, 0.45, 0.27}
\definecolor{forestgreen}{rgb}{0.13, 0.55, 0.13}

\usepackage{array}

\usepackage{tikz}
\usetikzlibrary{positioning}
\usetikzlibrary{decorations.pathreplacing,decorations.markings}
\usetikzlibrary{arrows.meta}

\definecolor{yellow2}{rgb}{0.98, 0.80, 0.20}
\definecolor{blue2}{RGB}{65,128,255}
\newcolumntype{C}[1]{>{\centering\let\newline\\\arraybackslash\hspace{0pt}}m{#1}}

\newcommand{\floor}[1]{ \left\lfloor{ #1} \right\rfloor}

\newcommand{\ov}{\over}

\newcommand{\beas}{\begin{equation} \begin{aligned}} \newcommand{\eeas}{\end{aligned} \end{equation}}

\newcommand{\Z}{\mathbb{Z}}
\newcommand{\C}{\mathbb{C}}

\title{On the Classification of Duality Webs for Graded Quivers} 
\author[a,b,c]{Sebasti\'an Franco,} 
\author[d]{Azeem Hasan,}
\author[e]{Xingyang Yu}

\affiliation[a]{
Physics Department, The City College of the CUNY \\
160 Convent Avenue, New York, NY 10031, USA}

\affiliation[b]{Physics Program and $^c$Initiative for the Theoretical Sciences \\
The Graduate School and University Center, The City University of New York  \\
365 Fifth Avenue, New York NY 10016, USA}

\affiliation[d]{Dipartimento di Matematica e Fisica Ennio De Giorgi, \\
Universit\`{a} del Salento \& INFN, Via Arnesano, 73100 Lecce, Italy}

\affiliation[e]{Center for Cosmology and Particle Physics, \\
New York University, 726 Broadway, New York, NY 10003, USA}

\emailAdd{sfranco@ccny.cuny.edu}
\emailAdd{azeem.hasan@le.infn.it} 
\emailAdd{xy1038@nyu.edu}

\abstract{We study the $m$-graded quiver theories associated to CY $(m+2)$-folds and their order $(m+1)$ dualities. We investigate how monodromies give rise to mutation invariants, which in turn can be formulated as Diophantine equations characterizing the space of dual theories associated to a given geometry. We discuss these ideas in general and illustrate them in the case of orbifold theories. Interestingly, we observe that even in this simple context the corresponding Diophantine equations may admit an infinite number of seeds for $m\geq 2$, which translates into an infinite number of disconnected duality webs. Finally, we comment on the possible generalization of duality cascades beyond $m=1$.
}

\begin{document}

\maketitle

%=================================================================
\section{Introduction}
%=================================================================

Minimally supersymmetric gauge theories in $6-2m$ dimensions enjoy order $(m+1)$ dualities, which generalize the celebrated $4d$ $\mathcal{N}=1$ Seiberg duality \cite{Seiberg:1994pq}. In particular, $2d$ $\mathcal{N}=(0,2)$ gauge theories exhibit {\it triality} \cite{Gadde:2013lxa}, while a {\it quadrality} has been proposed for $0d$ $\mathcal{N}=1$ gauge theories \cite{Franco:2016tcm}. A unified mathematical framework encompassing all such gauge theories in different dimensions and their dualities is provided by $m$-graded quivers with superpotentials \cite{Franco:2017lpa} (see also \cite{Closset:2017yte,Closset:2017xsc,Eager:2018oww} for related ideas). The cases of $m=0,1,2,3$ correspond to $6d$ $\mathcal{N}=(0,1)$, $4d$ $\mathcal{N}=1$, $2d$ $\mathcal{N}=(0,2)$ and $0d$ $\mathcal{N}=1$ field theories, respectively.

A large class of these gauge theories can be engineered on the worldvolume of Type IIB D$(5-2m)$-branes probing singular Calabi-Yau (CY) $(m+2)$-folds for $m\leq 3$. While this upper bound on $m$ is enforced by the critical dimension of Type IIB string theory, the $m$-graded quiver framework and their corresponding order $(m+1)$ mutations generalizing dualities extend to arbitrary $m$. These theories indeed have a physical interpretation as describing fractional branes at CY$_{m+2}$ singularities in the topological B-model \cite{Closset:2018axq}.

One of the primary goals of this paper is to classify and characterize the spaces of theories connected by these dualities. The (generically infinite) set of dual theories can be organized into so-called {\it duality webs}. Duality webs were first introduced for $m=1$ in \cite{Cachazo:2001sg}, with further studies appearing in \cite{Franco:2003ja}. Simple examples for $m=2$ and $m=3$ were investigated in \cite{Gadde:2013lxa, Franco:2016nwv} and \cite{Franco:2016tcm}, respectively.

A central question that we will address is whether it is possible to provide, for general $m$, a {\it global} characterization of the $m$-graded quiver theories connected by dualities, i.e. belonging to the same duality web. In other words, are there conditions/equations that the sets of theories related by dualities must satisfy? If so, is it possible to find these theories by solving these equations instead of explicitly acting with the mutations? For $m=1$, it is known that the realization of these theories in terms of D-branes and mirror symmetry leads to Diophantine equations that the corresponding $4d$ $\mathcal{N}=1$ theories must satisfy (see \cite{Cachazo:2001sg,Feng:2002kk,Franco:2002mu,Hanany:2012mb} for detailed discussions). These ideas are intimately related to solitons in $2d$ $\mathcal{N}=(2,2)$ theories \cite{Cecotti:1992rm,Cachazo:2001sg,Feng:2002kk}. In this paper we will extend to arbitrary $m$ the classification of dual theories via Diophantine equations, exploring the new features that arise from this generalization.

In their pioneering work \cite{Klebanov:2000hb}, Klebanov and Strassler introduced the concept of {\it duality cascade} for $4d$ $\mathcal{N}=1$ theories. This novel type of renormalization group (RG) flow takes the form of a sequence of Seiberg duality transformations in which a gauge group is dualized every time it goes to infinite coupling, switching its behavior from asymptotically free to IR free. Generically, the duality also modifies the scale dependence of other gauge couplings. 

In the context of duality cascades, duality webs become a chart of possible RG trajectories \cite{Cachazo:2001sg,Franco:2003ja}. Periodic cascades, in which the RG periodically alternates between a finite number of dual theories, are particularly elegant. In this case, the RG flow repeatedly goes around a closed cycle of the corresponding duality web. Remarkably, it was recently discovered that duality cascades describe topological transitions in certain $4d$ non-SUSY theory as parameters are varied \cite{Karasik:2019bxn}. It is natural to ask whether $(m+1)$-dualities also lead to duality cascades for $m>1$ and, if so, what their physical interpretation is. 

This paper is organized as follows. \sref{section_graded_quivers_and_mutations} reviews $m$-graded quivers and their dualities. \sref{section_monodromies} discusses monodromies, their connection to mirror symmetry and how they give rise to Diophantine equations. \sref{section_examples} illustrates these ideas for a family of orbifolds. \sref{section_C4Z4} investigates in more depth the Diophantine equations for $\mathbb{C}^4/\mathbb{Z}_4$ and their solutions. The possibility of duality cascades for general $m$ is discussed in \sref{section_cascades}. \sref{section_conclusions} summarizes our conclusions.

%=================================================================
\section{Graded Quivers and Mutations}
%=================================================================

\label{section_graded_quivers_and_mutations}

In this section we briefly review $m$-graded quivers and their order $(m+1)$ mutations, and explain their connections to physics. Towards the end, we discuss new simple consistency conditions that follow from the mutations. We refer the reader to \cite{Franco:2017lpa,Closset:2018axq} for detailed presentations and to \cite{lam2014calabi} for a mathematical analysis. Related works include \cite{Closset:2017yte,Closset:2017xsc, Eager:2018oww,Franco:2019bmx}.

%=================================================================
\subsection{Graded Quivers}
%=================================================================

\label{section_graded_quivers}

Given an integer $m \geq 0$, an $m$-graded quiver is a quiver equipped with a grading for every arrow $\Phi_{ij}$ by a {\it quiver degree}:
\beq
|\Phi_{ij}| \in \{ 0, 1, \cdots, m\}~.
\eeq
To every node $i$ we associate a unitary ``gauge group" $U(N_i)$. Arrows connecting nodes correspond to bifundamental or adjoint ``fields".\footnote{We will not consider theories with gauge groups that are not unitary or with fields that do not transform in the bifundamental or adjoint representations in this paper. The framework of $m$-graded theories can be extended to such non-quiver theories.}

For every arrow $\Phi_{ij}$, its conjugate has the opposite orientation and degree $m-|\Phi_{ij}|$:
\beq\label{Phi opp intro}
\overline{\Phi}_{ji}^{(m-c)}\equiv \overline{(\Phi_{ij}^{(c)})}~.
\eeq
Here we have introduced a notation in which the superindex explicitly indicates the degree of the corresponding arrow, namely $|\Phi_{ij}^{(c)}|=c$. 

The integer $m$ determines the possible degrees, therefore different values of $m$ give rise to qualitatively different classes of graded quivers. We can restrict the different types of arrows to have degrees in the range:
\beq\label{arrows fields}
\Phi_{ij}^{(c)} \; : i \longrightarrow j~, \qquad c=0, 1, \cdots, n_c-1~, \qquad n_c \equiv    \floor{m+2\over 2}~,
\eeq
since other degrees can be obtained by conjugation.\footnote{The range in \eref{arrows fields} is just a conventional choice. The $n_c$ ``fundamental" degrees can be picked differently. Sometimes it is convenient to deal with all possible values of the degrees. For every arrow, either $\Phi_{ij}^{(c)}$ or $\overline{\Phi}_{ji}^{(m-c)}$ can be regarded as the fundamental object.} We refer to degree 0 fields as {\it chiral fields}.

Graded quivers for $m=0,1,2,3$ describe $d=6,4,2,0$ minimally supersymmetric gauge theories, respectively. Different degrees translate into different types of superfields. Table \eref{table_graded_quivers_QFTs} summarizes the correspondence between graded quivers and gauge theories. We also indicate how some of these theories can be engineered in terms of Type IIB D$(5-2m)$-branes probing CY $(m+2)$-folds.
\be
\begin{tabular}{c|cccc}
$m$ & $0$ &$1$& $2$ & $3$  \\
 \hline
CY    &CY$_2$ &CY$_3$ & CY$_4$& CY$_5$\\
SUSY & $6d$  $\CN=(0,1)$ & $4d$ $\CN=1$ & $2d$ $\CN=(0,2)$ & $0d$ $\CN=1$ 
\end{tabular}
\label{table_graded_quivers_QFTs}
\ee

%=================================================================       
\paragraph{Superpotential.}
%=================================================================       
The {\it superpotential} of an $m$-graded quiver consists of a linear combination of gauge invariant terms, i.e. closed oriented cycles in the quiver, of degree $m-1$:
\be
W= W(\Phi)~,\qquad\qquad  |W|= m-1~.
\label{superpotential_degree}
\ee

There is no possible superpotential for $m=0$. For $m=1,2,3$, the superpotentials take the schematic forms:
\beq
\begin{array}{ll}
m=1 : \ &W= W(\Phi^{(0)})~, \\[.35cm]
m=2 : \  &W= \Phi^{(1)}J(\Phi^{(0)})+ \overline{\Phi}^{(1)} E(\Phi^{(0)})~, \\[.35cm]
m=3 : \  &W= \Phi^{(1)}\Phi^{(1)} H(\Phi^{(0)})+ \Phi^{(2)} J(\Phi^{(0)})~, 
\end{array}
\label{W_different_m}
\eeq
where $W(\Phi^{(0)})$, $J(\Phi^{(0)}), E(\Phi^{(0)})$ and $H(\Phi^{(0)})$ are holomorphic functions of the chiral fields.

%=================================================================       
\paragraph{Kontsevich bracket condition.}
%================================================================= 
In addition to the constraint on its degree \eref{superpotential_degree}, the superpotential must also satisfy:
\beq
\{W,W \}=0 ~.
\label{superpotential_Kontsevich}
\eeq 
Here $\{ f, g \}$ denotes the Kontsevich bracket, which is a natural generalization of the Poisson bracket to a graded quiver (see e.g. \cite{Franco:2017lpa,Closset:2018axq} for details).

%=================================================================       
\subsection{Generalized anomaly cancellation}
 %=================================================================  

Graded quivers must also satisfy the {\it generalized anomaly cancellation} conditions \cite{Franco:2017lpa}. For $m$ odd, these conditions are given by:
\beq
\sum_j N_j \sum_{c=0}^{n_c-1} (-1)^c \left(\CN(\Phi_{ji}^{(c)})-\CN(\Phi_{ij}^{(c)})\right)=0~, \qquad \forall i~, \qquad {\rm if}\;\; m \in 2\mathbb{Z}+1~,
\label{anomaly_odd}
\ee
where $N_j$ is the rank of the $j^{th}$ node and $\CN(\Phi_{ij}^{(c)})$ denotes the number of arrows from $i$ to $j$ of degree $c$. For every fixed $i$, the sum over $j$ runs over all nodes in the quiver (including $i$), and $n_c$ is given by \eref{arrows fields}. For $m$ even, the conditions become
\be
\sum_j N_j \sum_{c=0}^{n_c-1}(-1)^c \left(\CN(\Phi_{ji}^{(c)})+\CN(\Phi_{ij}^{(c)})\right)=2N_i~, \qquad \forall i~, \qquad {\rm if}\;\; m \in 2\mathbb{Z}~.
\label{anomaly_even}
\ee
These conditions follow from requiring the invariance of the ranks under $(m+1)$ consecutive mutations. For $m=0,1,2,3$, they correspond to the cancellation of non-abelian anomalies for the corresponding $d=6,4,2,0$ gauge theories with gauge group $\prod_i U(N_i)$.

%=================================================================       
\subsection{Mutations}
 %=================================================================  
 
$m$-graded quivers admit order $(m+1)$ mutations.\footnote{For brevity, throughout this paper we will use mutation and dualization/duality interchangeably.} For $m\leq 3$, these mutations reproduce the dualities of the corresponding gauge theories, namely: no duality for $6d$ $\mathcal{N}=(0,1)$, Seiberg duality for $4d$ $\mathcal{N}=1$ \cite{Seiberg:1994pq}, triality for $2d$ $\mathcal{N}=(0,2)$ \cite{Gadde:2013lxa} and quadrality for $0d$ $\mathcal{N}=1$ \cite{Franco:2016tcm}. Moreover, the mutations provide a generalization of these dualities to $m>3$. It is natural to expect that these generalized dualities correspond to mutations of exceptional collections of B-branes in CY $(m+2)$-folds. Since this paper focuses on the space of theories connected by such mutations, we briefly review them in this section for completeness. We refer to \cite{Franco:2017lpa} for further details. We now summarize the effect of a mutation on a node of the quiver, which we denote by $\star$.

\begin{figure}[H]
	\centering
	\includegraphics[width=12.5cm]{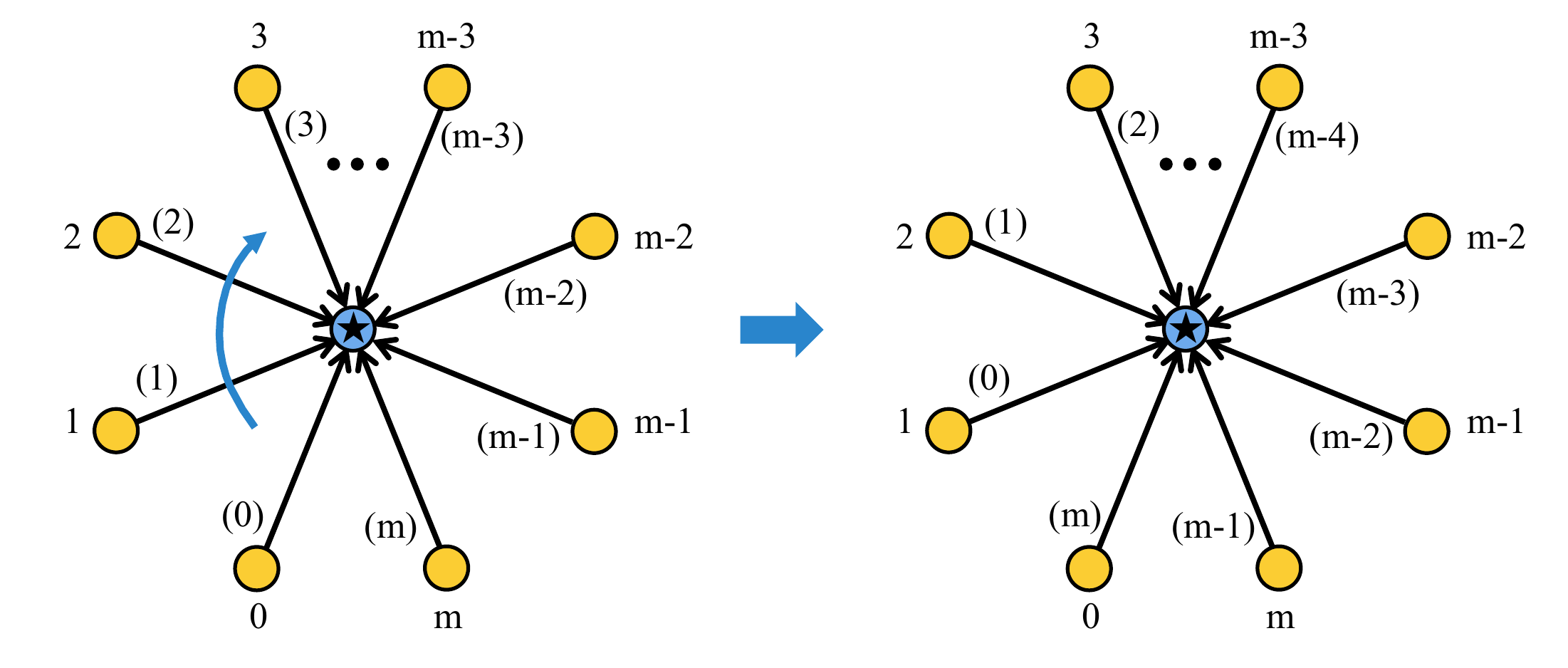}
\caption{The transformation of flavors upon a mutation on node $\star$ can be implemented as a rotation of the degrees of the arrows.}
	\label{mutation_flavors}
\end{figure}

%=================================================================
\paragraph{1. Flavors.}
%=================================================================
We refer to the arrows connected to the mutated node as {\it flavors}. We can take all flavors as incoming into the mutated note, simply by trading outward oriented arrows for their conjugate. After doing this, there is a natural cyclic order for flavors around the node, in which the degree of incoming arrows increases clockwise, as shown on the left of \fref{mutation_flavors}. There can be multiple arrows of a given degree.\footnote{In \sref{section_additional_conditions_mutations} we will discuss whether it is possible for arrows of some degree to be absent, by considering additional consistency constraints for graded quivers that arise from mutations.}

%=================================================================
%=================================================================

Under the mutation, the flavors transform as follows:
\paragraph{2. Rotation of the degrees.} {Replace every incoming arrow} \xymatrix{ i \ar@{->}[r]^{(c)} & \star} with the arrow 
\xymatrix{ i \ar@{->}[r]^{(c-1)} & \star}.
In terms of the cyclic ordering of flavors, this transformation is implemented as a clockwise rotation of the degrees of the flavors while keeping the spectator nodes fixed, as shown in \fref{mutation_flavors}.

%=================================================================
\paragraph{2. Mesons.} 
%=================================================================
Next we add composite arrows, to which we refer as {\it mesons}. For every $2$-path \xymatrix{ i \ar@{->}[r]^{(0)} & \star \ar@{->}[r]^{(c)} & j} in the quiver, where $c\neq m$, {we add a new arrow} \xymatrix{ i  \ar@/^0.5pc/[rr]^{(c)} & \star & j}. In summary, we generate all possible mesons involving incoming chiral fields. Sometimes, we might represent the field to be composed with a chiral field as an incoming arrow into the mutated node. The orientations of both incoming arrows naively seem incompatible for composition. The general rule above means that, in such cases, we use the conjugate of the incoming chiral field for the composition. This phenomenon, denoted {\it anticomposition}, was first discussed in the physics literature in the context of quadrality of $0d$ $\mathcal{N}=1$ theories \cite{Franco:2016tcm}. 
%=================================================================
\begin{figure}[H]
	\centering
	\includegraphics[width=12cm]{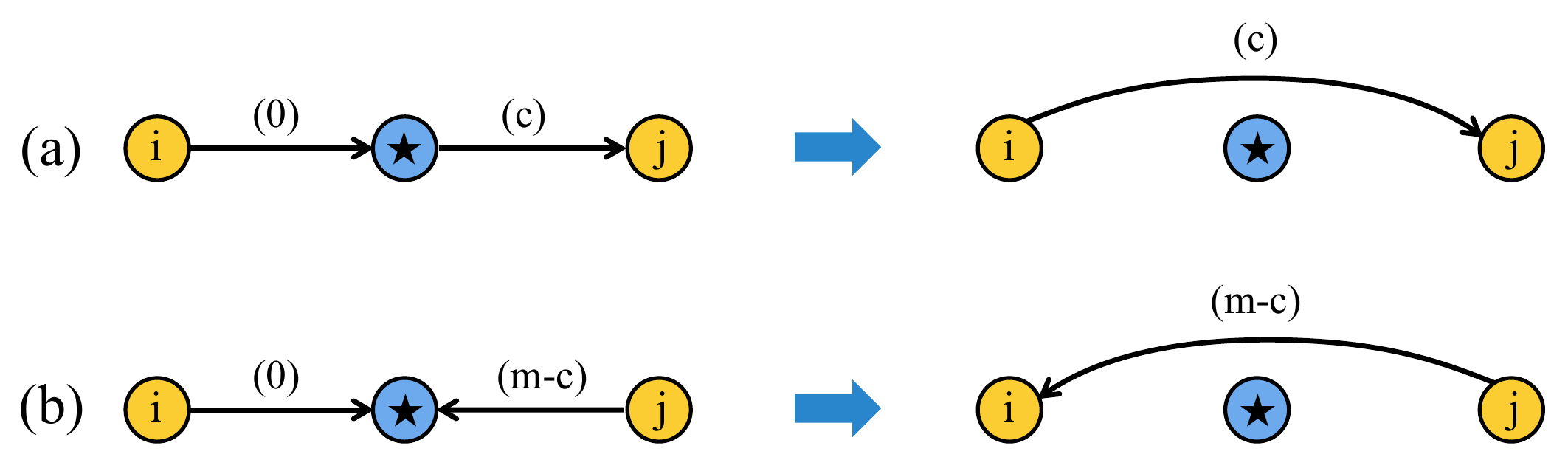}
\caption{a) Composition of arrows into a meson. b) The same process interpreted as anticomposition.}
	\label{composition_anticomposition}
\end{figure}
%=================================================================

%=================================================================
\paragraph{3. Superpotential.} 
%=================================================================

The superpotential transforms according to four rules. For brevity, we will not review them here and instead refer the reader to \cite{Franco:2017lpa} where they were originally presented.

%=================================================================
\paragraph{4. Ranks.} 
%=================================================================
The rank $N_\star$ of the mutated node transforms as:
\beq
N_\star'=N_0-N_\star~,
\label{mutation_ranks}
\eeq
where $N_0$ indicates the total number of incoming chiral fields.
Periodicity of the rank after $(m+1)$ consecutive mutations on the same node implies the anomaly cancellation conditions \eref{anomaly_odd} and \eref{anomaly_even}.

%=================================================================       
\subsection{Additional Consistency Conditions from Mutations}
%=================================================================  
 
 \label{section_additional_conditions_mutations}

Mutations give rise to additional consistency conditions for graded quivers. Let us focus on nodes that do not contain adjoint fields, since the mutation we have discussed only applies to this case. Whether nodes with adjoints can be mutated and, if so, how are interesting questions that deserve further study. Denoting $N^{(c)}_i$ the number of fields of degree $c$ in the fundamental representation of node $i$, we must have
\beq
N^{(0)}_i \geq N _i ~.
\label{SUSY_breaking_condition}
\eeq
If instead $N^{(0)}_i < N_i$, a mutation on node $i$ would turn its rank negative. This is the analogue, for arbitrary $m$, of the $N_f < N_c$ regime in $4d$ SQCD. For $m\leq 3$, it is natural to conjecture that this regime is related to SUSY breaking. Similarly, $N^{(0)}_i = N _i$ is analogous to $N_f = N_c$ for $4d$ SQCD. In this case, formal application of the mutation leads to $N_i'=0$, i.e. to the disappearance of node $i$. In the case of theories associated to toric CY $(m+2)$-folds, this condition was proposed as the one for reducibility of the corresponding $m$-dimers \cite{Franco:2018qsc}.

The mutation rule for flavors summarized in \fref{mutation_flavors} implies that incoming flavors of degree $c$ become incoming chirals after $c$ consecutive mutations on node $i$. Then, dualizing node $i$ multiple times, \eref{SUSY_breaking_condition} generalizes to
\beq
\begin{array}{ccl}
N^{(1)}_i & \geq & N^{(0)}_i-N _i \\[.1cm]
N^{(2)}_i & \geq & N^{(1)}_i-N^{(0)}_i+N _i \\
\vdots & & \vdots \\
N^{(m)}_i & \geq & N^{(m-1)}_i-N^{(m-2)}_i+ \cdots \pm N _i 
\end{array}
\label{conditions_from_mutations}
\eeq
where the RHS are the ranks of the gauge group after different number of mutations. The inequalities must be strict if we want to avoid reducibility. In particular, \eref{conditions_from_mutations} implies that not only the quiver, but every node must contain flavors of all possible degrees. 

While random quivers generically violate these conditions, it is worth emphasizing that all the explicit examples of quivers for branes on CY $(m+2)$-folds satisfy \eref{conditions_from_mutations} (see e.g. \cite{Franco:2015tna,Franco:2015tya,Franco:2016nwv,Franco:2016qxh,Franco:2016fxm,Franco:2016tcm,Franco:2017cjj,Franco:2018qsc,Closset:2018axq,Franco:2019bmx}).

%=================================================================
\section{Monodromies, Diophantine Equations and Mirror Symmetry}
%=================================================================

\label{section_monodromies}

It is well known that the soliton spectrum of $2d$ $\mathcal{N}=(2,2)$ is related to the intersections of vanishing cycles at singularities, and hence to the corresponding quivers \cite{Cecotti:1992rm,Cachazo:2001sg,Feng:2002kk}. This connection arises naturally in the context of mirror symmetry and has been explored in detail in the case of $4d$ $\mathcal{N}=1$ quiver theories for CY 3-folds \cite{Cachazo:2001sg,Feng:2002kk}. In this section, we discuss how this correspondence applies to CY$_{m+2}$ folds, emphasizing some of the specific features related to $m$-graded quivers. In particular, we explain how monodromies lead to invariants under order $(m+1)$ mutations, which can be expressed as sets of Diophantine equations constraining the field content of the quivers.

%=================================================================
\subsection{Monodromy from $2d$ $\mathcal{N}=(2,2)$ Supersymmetry} 
%=================================================================

\label{section_monodromy_from_SUSY}

We first briefly review some of the results in Cecotti and Vafa's seminal work \cite{Cecotti:1992rm}, which classified the vacuum structure of $2d$ $\mathcal{N}=(2,2)$ theories using singularity theory. Consider a $2d$ $\mathcal{N}=(2,2)$ theory consisting of chiral fields $P(x_{\mu})$,  $\mu=1,\ldots,m+1$, and with a superpotential $W$ equal to the Newton polynomial of a toric CY $(m+2)$-fold, namely
\beq
W=P(x_1,\ldots, x_{m+1})=\sum_{\vec{v} \in V} c_{\vec{v}} \, x_1^{v_1} \ldots x_{m+1}^{v_{m_1}} , 
\label{Newton_polynomial}
\eeq
where the $c_{\vec{v}}$ are complex coefficients and the sum runs over points $\vec{v}$ in the toric diagram $V$. It is possible to scale $m+2$ of the coefficients to 1.

A vacuum of this theory corresponds to a critical point $x^{(i)}_*\equiv (x^{(i)}_{*,1},\ldots,x^{(i)}_{*,m+1})$ of $W$, namely
\beq
\left. {\partial \over \partial x_\mu} W \right |_{(x^{(i)}_{*,1},\ldots,x^{(i)}_{*,m+1})}=0 \ \ \ \ \forall \, \mu .
\label{critical_points}
\eeq
where $i=1,\ldots,n_*$ labels the critical point. We assume that the superpotential is quadratic and non-degenerate around every critical point. This can be arranged by perturbing the superpotential slightly, if necessary.

If the toric diagram $V$ has at least one internal point, then the number of critical points is 
\beq
n_* = (m+1)!\, \mbox{Vol}(V) \, ,
\eeq
as shown in \cite{Feng:2005gw}. Equivalently, $n_*$ is the volume of the toric diagram normalized such that the smallest $(m+1)$-dimensional lattice simplex has volume $1$. 

Let us now consider solitons, i.e. field configurations connecting distinct vacua. We can construct a basis of solitons by picking a non-critical point $t$ and considering vanishing cycles along the vanishing path $\gamma_{i}$, which is a straight segment connecting $P(t)$ to $P(x_*{(i)})$. Every intersection between these vanishing cycles corresponds to a soliton between the corresponding vacua. Perturbing the superpotential moves the image of the critical points $P(x_i^*)$ on the $P$-plane. The intersection between two vanishing cycles can only change when the two $\gamma$'s pass through each other. The choice of $t$ induces a cyclic ordering of the images $P(x_*{(i)})$ around $P(t)$. Let us consider what happens when $\gamma_{i}$ is moved over the adjacent path $\gamma_{i+1}$. In this case, the net intersection number $\mu_{i,j}$ of solitons connecting vacua $i$ and $j$, counted with orientation, changes as follows
    \begin{align}
         \mu_{i+1,j} &\to \mu_{i,j} \nonumber \\
         \mu_{i,j} &\to \mu_{i,i+1}\mu_{i+1,j} - \mu_{i,j} \label{cycle_crossing}
    \end{align}  
In the above expression we require $j > i$ using the cyclic ordering induced by the non-critical point $t$ and an arbitrary reference vanishing cycle.

The resulting monodromy matrix $M$ can be expressed in terms of $\mu_{i,j}$ by defining an upper triangular matrix $S$ as follows
\beq
        S_{ij} = \left\{\begin{matrix}
                         1 && i = j \\
                         \mu_{i,j} && j > i \\
                         0 && i < j
                       \end{matrix} 
                 \right. 
\eeq
In terms of $S$, $M$ is simply given by
\beq
M = S^{-T}S \, .
\eeq

The eigenvalues of $M$ are phases and remain unchanged under the transformations in \eref{cycle_crossing}. This results in important invariants of the geometry of solitons. Since the eigenvalues of $M$ are the roots of the characteristic polynomial 
\beq
Q(z) = \det(z-M) \, ,
\eeq
we conclude that $Q(z)$ is also invariant. Since $M$ is an integer matrix, this condition gives rise to Diophantine equations that the intersection numbers $\mu_{i,j}$ must satisfy. There is one such equation for every power of $z$ in the expansion of $Q(z)$. However, since $Q(z)$ satisfies 
\beq
Q(z^{-1}) = \pm z^{-n_*}Q(z) \, ,  
\eeq 
not all the coefficients in this expansion are independent. The coefficients of the $z^{i}$ and $z^{n_*-i}$ terms are equal, so we obtain only $\lceil n_*/ 2\rceil$ independent equations.

%=================================================================
\subsection{$m$-Dimers in the Mirror}
%=================================================================

\label{subsection:dimers_mirror}

The classification of solitons described in the previous section can be mapped to the construction of the mirror of the underlying toric CY $(m+2)$-fold. Succinctly, given a toric CY$_{m+2}$ $\mathcal{M}$ specified by a toric diagram $V$, the mirror geometry \cite{Hori:2000kt,Hori:2000ck} is an $(m+2)$-fold $\mathcal{W}$ defined as a double fibration over the complex $W$-plane
\beq
\begin{array}{rl}
W = & P(x_1,\ldots, x_{m+1}) \\[.1cm]
W = & uv
\end{array}
\label{double_fibration}
\eeq 
where $u,v \in \mathbb {C}$ and $x_\mu\in \mathbb{C^*}$, $\mu=1,\ldots,m+1$. $P(x_1,\ldots, x_{m+1})$ is the Newton polynomial defined in \eref{Newton_polynomial}. The critical points are given by \eref{critical_points} and the corresponding critical values on the $W$-plane are $W^{(i)}_*=P(x^{(i)}_{*,1},\ldots,x^{(i)}_{*,m+1})$.

The double fibration consists of a holomorphic $m$-complex dimensional surface $\Sigma_W$ coming from $P(x_1,\ldots, x_{m+1})$ and a $\mathbb{C}^*$ fibration from $uv$. The corresponding $S^{m}\times S^1$ is fibered over a vanishing path connecting $W=0$ and $W=W^{(i)}_*$, and gives rise to an $S^{m+2}$. We refer to these spheres as $\mathcal{C}_i$, $i=1,\ldots, n_*$. The $\mathcal{C}_i$ are in one-to-one correspondence with vanishing cycles $C_i$ at $W=0$, where the $S^1$ fiber vanishes. Every $\mathcal{C}_i$ gives rise to a {\it vanishing cycle} $C_i$ with $S^{m+1}$ topology. The $C_i$ live on the Riemann surface $\Sigma_0$, defined by $P(x_1,\ldots, x_{m+1})=0$. 

Every $C_i$ gives rise to a gauge group and the matter fields in the quiver correspond to their intersections. In fact, it is possible to use mirror symmetry to construct the full $m$-dimer, which encodes both the $m$-graded quiver and its superpotential. This construction has been elaborated in \cite{Feng:2005gw,Futaki:2014mpa,Franco:2016qxh,Franco:2016tcm,Franco:2017lpa}.

%=================================================================
\subsubsection{Grading}
%=================================================================

\label{section_mirror_grading}

The grading of fields on the quiver, or equivalently on the corresponding $m$-dimer, is not immediately manifest in the discussion of solitons in $2d$ $\mathcal{N}=(2,2)$ theories. How to directly determine the grading from the mirror is known for $m=1$ and $2$ \cite{Feng:2005gw,Franco:2016qxh}. While a general prescription for doing so for arbitrary $m$ is not yet known, this is not a problem since efficient alternative procedures for determining the degrees exist. In this section we discuss some of the important implication of grading.

%=================================================================
\paragraph{Ordering.}
%=================================================================

Consider an arbitrary vanishing cycle $C_\star$. Other vanishing cycles that intersect with $C_\star$ give rise to arrows connected to the corresponding node. The corresponding vanishing paths are cyclically ordered on the $W$-plane according to increasing degree of the fields associated with their intersections with $\gamma_\star$, conventionally oriented into the corresponding node.\footnote{An alternate way of obtaining the ordering of nodes is through an exceptional collection of sheaves.} This order is intimately related to the geometric realization of dualities as geometric transitions in the mirror \cite{Cachazo:2001sg,Franco:2016qxh,Franco:2016tcm}. 

%=================================================================
\begin{figure}[ht]
	\centering
	\includegraphics[width=7cm]{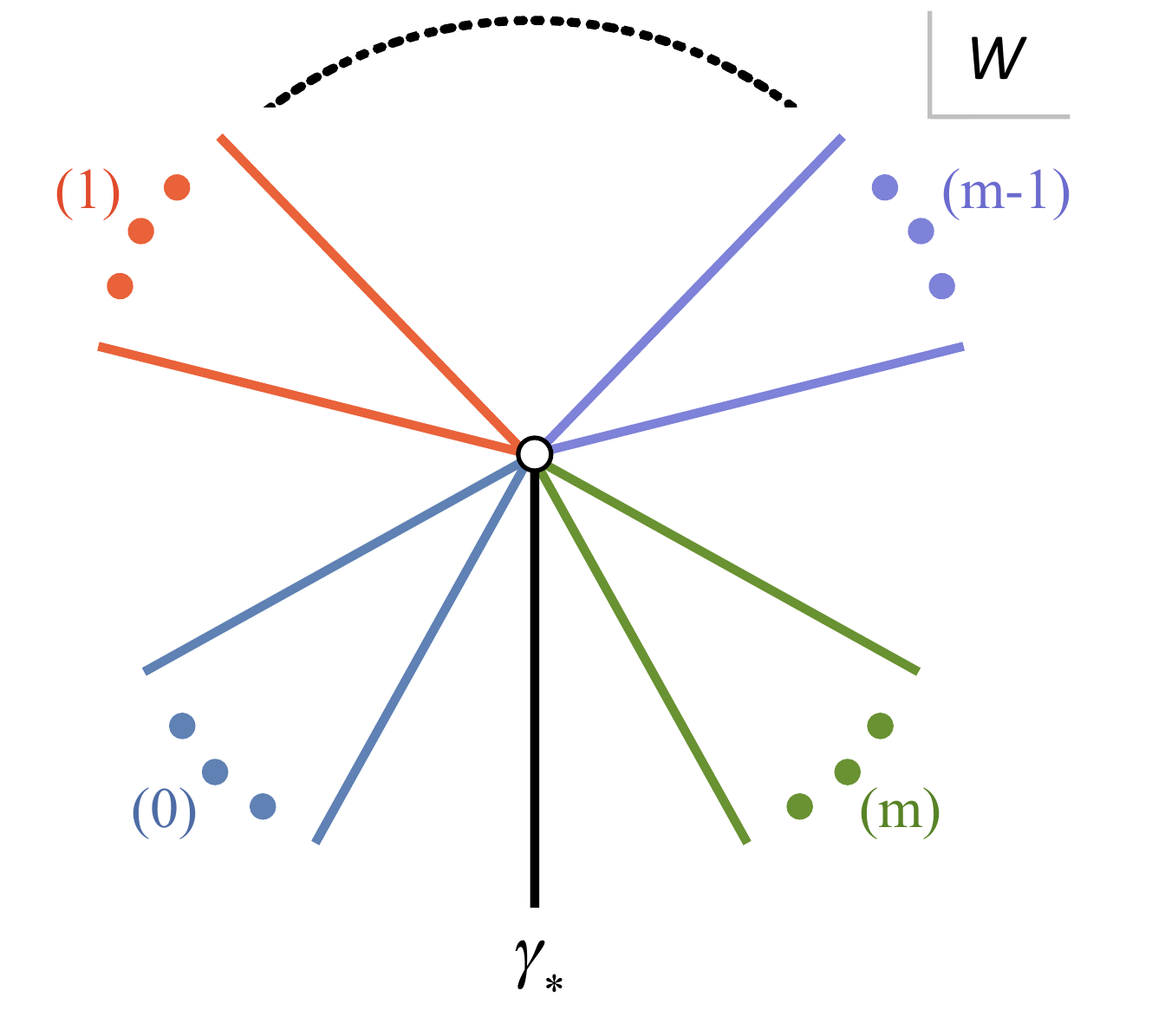}
\caption{For any reference cycle, the other vanishing paths are ordered on the $W$-plane according to the degree (indicated in parentheses) of their intersections with it.}
	\label{colors_mirror_W_plane}
\end{figure}
%=================================================================

Using this ordering, we can define the upper triangular matrix $S$ as
        \begin{align}
            S_{ij} = \left\{\begin{matrix}
                         1 && i = j \\
                         \sum_{c}(-1)^{c+1}n^{(c)}_{ij}  && j > i \\
                         0 && i < j
                       \end{matrix} 
                 \right. \label{S_matrix_quiver}
        \end{align}
where $n^{(c)}_{ij}$ is the number of fields of degree $c$ going from node $i$ to $j$. In terms of $S$ the monodromy matrix $M$ is again just $M = S^{-T}S$.

As explained in \sref{section_monodromy_from_SUSY}, the invariance of the characteristic polynomial $Q(z) = \det(z - M)$ under geometric transitions that reorder the vanishing cycles gives rise to $\lceil n_*/ 2\rceil$ Diophantine equations. These equations are satisfied by the quivers of every dual phase corresponding to a given underlying geometry.\footnote{As we elaborate below, the Diophantine equations are not only satisfied by the dual theories but also by theories related by more general transitions.}

%=================================================================
\paragraph{Dualities.}
%=================================================================

\begin{figure}[H]
	\centering
	\includegraphics[width=7cm]{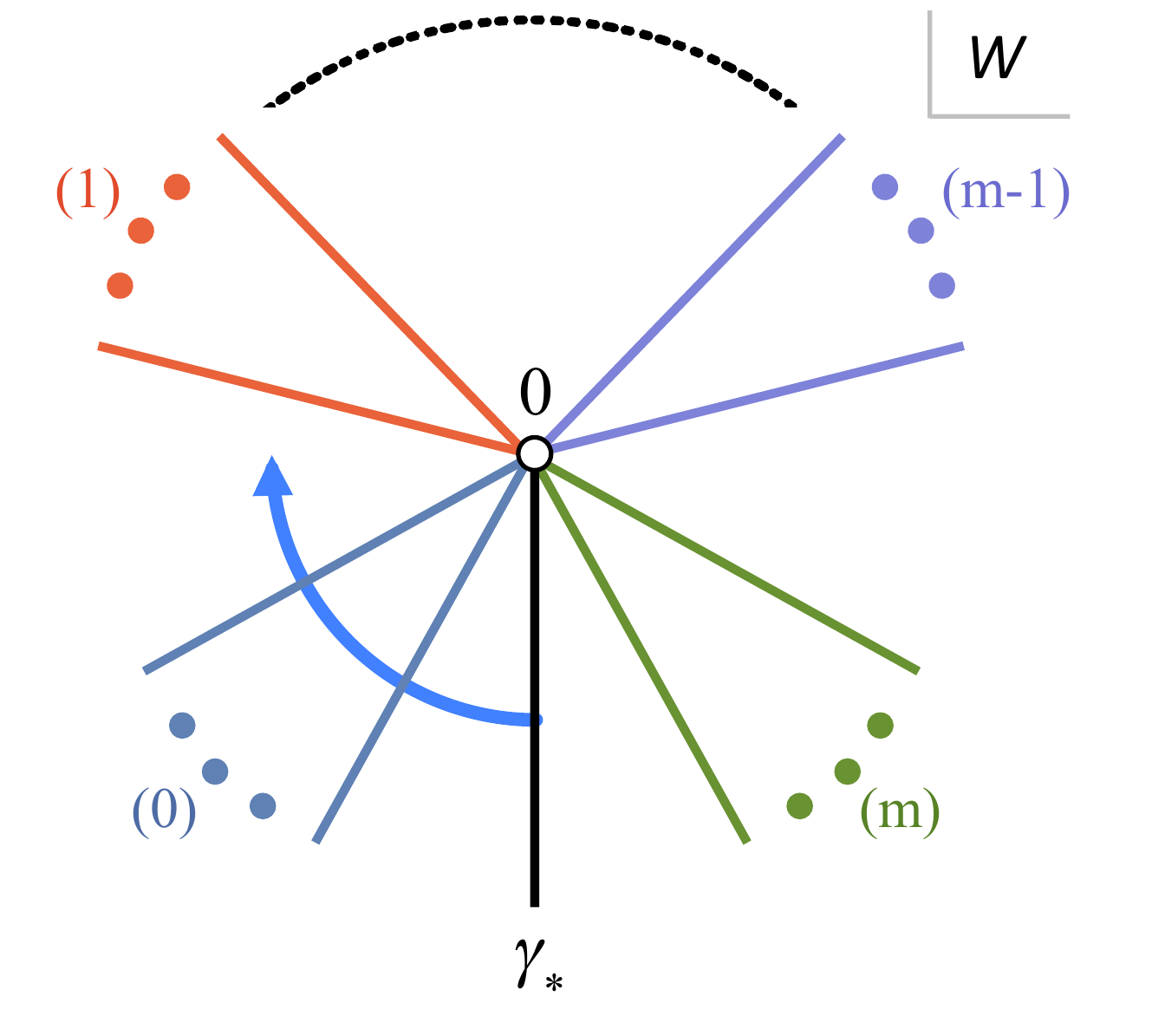}
\caption{Duality as a geometric transition in the mirror.}
	\label{mutation_mirror}
\end{figure}

The discussion in \sref{section_monodromy_from_SUSY} applies to arbitrary reordering of the vanishing cycles. It is a purely geometric statement, without any reference to the grading. However, grading plays a crucial role in determining the geometric transitions that correspond to the order $(m+1)$ dualities. For a given $m$, the transition associated to the corresponding duality is schematically shown in \fref{mutation_mirror} \cite{Franco:2017lpa}.

 %=================================================================
%=================================================================

It corresponds to moving the vanishing cycle associated to the mutated node past all the vanishing cycles contributing incoming chiral fields to it. In other words, while grading is not reflected in the monodromy matrix or the resulting Diophantine equations, it restricts the transitions associated to dualities. In what follows, we will focus on such transitions instead of generic ones. For $m=1$, generic transitions were studied in \cite{Feng:2002kk}, where they were referred to as {\it fractional Seiberg dualities}. Extending this nomenclature to any $m$, we denote the transitions that do not correspond to dualities as fractional dualities.
Fractional dualities correspond to reordering of vanishing cycles that violate the cyclic ordering of \fref{colors_mirror_W_plane}. It is important to emphasize that fractional duals still obey the Diophantine equations.

%=================================================================
\paragraph{Anomalies and other constraints.}
%=================================================================

It is natural to ask whether a given $S$ and the resulting $M$ are consistent with different gradings. More generally, one can also ask whether they can correspond to different values of $m$. If the latter was possible, fractional duals for a given $m$ could be duals for another $m'$. While we do not have a general answer to these questions, we note that it is a highly constrained problem. As mentioned in \sref{section_additional_conditions_mutations}, mutations give rise to multiple constraints, as shown in \eref{conditions_from_mutations}. 

In addition the sign of some of the eigenvalues of $M$ is related to the parity of $m$. The ranks $N_{i}$ of the gauge groups must satisfy the anomaly cancellation condition \eqref{anomaly_odd} or \eqref{anomaly_even}, which can be conveniently recast using the matrix $S$. The requirement that the vector of ranks $N$ satisfies anomaly cancellation means that
\beq
S \, N + (-1)^{m}S^{T} N = 0 \, .     
\eeq
Multiplying this equation by $S^{-T}$ on the left, we get
\beq
M \, N = (-1)^{m+1} N \, .
\eeq
Therefore, $N$ is an eigenvector of the monodromy matrix with eigenvalue $\pm 1$, where the sign depends on the parity of $m$. As already mentioned, the eigenvalues of $M$ are phases, but we see that the anomaly cancellation condition gives us a stronger constraint: at least one of them must be $1$ if $m$ is odd and $-1$ if $m$ is even.

It is possible that additional restrictions on the interplay between $S$ and grading, not mentioned in this section, exist.

%=================================================================
\section{Examples: $\mathbb{C}^{m+2}/\mathbb{Z}_{m+2}$}
%=================================================================

\label{section_examples}

In this section, we present explicit examples illustrating the classification of dual $m$-graded quivers associated to CY $(m+2)$-folds via Diophantine equations. In particular, we will focus in an infinite class of  
$\mathbb{C}^{m+2}/\mathbb{Z}_{m+2}$ orbifolds.

%=================================================================
\subsection{Geometry and Quiver Theories}
%=================================================================

\label{section_quivers_orbifolds}
   
Let us consider the orbifolds $\C^{m+2}/\Z_{m+2}$ for which the cyclic group acts on flat space as:
\be
z_i \sim e^{2\pi i \ov m+2} z_i~, \qquad i=1, \cdots, m+2~,\quad \qquad (z_i) \in \C^{m+2}\, .
\ee
These singularities can be resolved to local $\mathbb{P}^{m+1}$.    

Let us now discuss the corresponding quiver theories. The $m=0$ and $1$ cases have been thoroughly studied in the literature. For early references on $m=2,3,4$, see \cite{GarciaCompean:1998kh,Franco:2015tna,Franco:2016tcm,Franco:2017lpa}. The quiver theories for arbitrary $m$ were first presented in \cite{Closset:2018axq}, where they were independently derived using both a combination of dimensional reduction and orbifolding, and the topological B-model. Here we briefly review some of the key results.

The toric diagram for $\mathbb{C}^{m+2}/\mathbb{Z}_{m+2}$ consists of the following $m+3$ points:
            \begin{align}
                v_0=(0,\ldots,0)~, 
                \qquad 
                \begin{array}{l}
                v_1= (1,0,0,\ldots,0)~, \\
                 v_2=(0,1,0,\ldots,0)~, \\
                \quad \vdots \\
                 v_{m+1}=(0,0,\ldots,0, 1)~, 
                \end{array} 
                \qquad 
                v_{m+2}=(-1,-1, \ldots, -1)~.
                 \label{eq_toric_Cn_Zn}
            \end{align} 
Of these, $v_{0}$ is an internal point while the rest are the vertices of an $(m+1)$-simplex. 
            
The quiver has $m+2$ gauge groups, which we label by the integers $0, \ldots, m+1$. This fact follows from the order of the orbifold group and is also reflected in the normalized volume of the toric diagram \eref{eq_toric_Cn_Zn}. For every $m$ there is a single toric phase, i.e. a phase described by an $m$-dimer. For such phase, all the gauge groups have the same rank and the matter content can be summarized as 
\beq
                \Phi_{i,i+k}^{(k-1;\mathbf{a}_{k})} : i \xrightarrow[(k-1)]{\,\,\,\,\,\,\,\, \binom{m+2}{k}\,\,\,\,\,\,\,\,} i+k \qquad 0 \le i < m+2 ; 1 \le k < m+2 - i \, ,
\eeq
where the notation is as follows: $\Phi_{i,j}^{(c;\mathbf{r})}$ represents a multiplet in the bifundamental representation of the $i$ and $i+j$ gauge groups, which has degree $c$ and transforms in the representation $\mathbf{r}$ of the global $SU(m+2)$ symmetry. The totally antisymmetric $k$-index representation is indicated as $\mathbf{a}_{k}$. The numbers above and below the arrow indicate the multiplicity and degree, respectively.

The superpotential is cubic and can be succinctly written as 
\beq
                W = \sum_{i+j+k < m+2}\Phi_{i,i+j}^{(j-1;\mathbf{a}_{j})}\Phi_{i+j,i+j+k}^{(k-1;\mathbf{a}_{k})}\bar{\Phi}_{i+j+k,i}^{(m+1-j-k;\mathbf{a}_{m+2-j-k})} ~.
                \label{potential_Cn_Zn_toric}
\eeq
Every term has $m+2$ $SU(m+2)$ indices. We have suppressed these indices and implicitly contracted them with a Levi-Civita tensor to form an $SU(m+2)$ singlet.

%=================================================================    
\subsection{General Structure of the Duals} 
%=================================================================
  
We now prove by induction general properties of the dual phases of $\mathbb{C}^{m+2}/\mathbb{Z}_{m+2}$, namely of the theories connected by an arbitrary sequence of mutations to the toric phases presented in the previous section. In particular, we will show that all these theories have cubic superpotential and are {\it monochromatic}. Monochromaticity has been introduced in the mathematical literature to indicate quivers in which all the fields connecting any pair of nodes, considering orientation, have the same degree \cite{MR2553375}. Interestingly, acting with mutations on a monochromatic quiver does not generate any adjoint. However, it is important to note that, in general, monochromaticity is not preserved under mutation, since mesons stretching between two nodes may have a different degree than the preexisting fields connecting them. When this occurs, monochromaticity might be restored if the appropriate fields become massive. We will explicitly verify this property for the theories under consideration.\footnote{This property has a nice characterization in the B-model realization of $m$-dimers in terms of exceptional collections of sheaves. Under suitable conditions, an exceptional collection gives rise to a monochromatic quiver. The requirement that the quiver remains monochromatic after an arbitrary sequence of mutations gives rise to the stronger constraint that the exceptional collection is part of a helix \cite{Herzog:2004qw,Herzog:2005sy,Herzog:2006bu,MR1074776}.}

Let us consider an arbitrary phase of $\mathbb{C}^{m+2}/\mathbb{Z}_{m+2}$. The bifundamental fields connecting nodes $i$ and $j$ transform in a general representation $\mathbf{r}_{i,j}$ of $SU(m+2)$. Let us assume that the superpotential of this theory is cubic, i.e. that it takes the form
\beq
W = \sum_{i+j+k < m+2}\Phi_{i,i+j}^{(j-1;\mathbf{r}_{i,i+j})}\Phi_{i+j,i+j+k}^{(k-1;\mathbf{r}_{i,i+k})}\bar{\Phi}_{i+j+k,i}^{(m+1-j-k;\bar{\mathbf{r}}_{i+j+k,i})} ~,
\label{potential_Cn_Zn_general}
\eeq
where $\bar{\mathbf{r}}_{j,i}$ is the conjugate representation of $\mathbf{r}_{i,j}$. In writing this expression, we have assumed that we can label the nodes such that the degree of fields between nodes $i$ and $j$ is $j-i-1$. Furthermore, \eref{potential_Cn_Zn_general} is schematic and should be understood as the $SU(m+2)$ singlet resulting from the combination of these fields. The existence of such a singlet imposes 
constraints on the representations $\mathbf{r}_{i,j}$. In this particular case, the following fusion rule holds
\beq
            \mathbf{r}_{ij} \otimes \mathbf{r}_{jk} \supseteq \mathbf{r}_{ik} \qquad i < j < k   
            \label{cn_zn_rep_fusion}
\eeq
which in turns implies
\beq
            \mathbf{r}_{ij} \otimes \mathbf{r}_{jk} \otimes \bar{\mathbf{r}}_{ki} \supseteq \mathbf{r}_{ik} \otimes \bar{\mathbf{r}}_{ki} \supseteq \mathbf{1} ~,
\eeq
which is the singlet in the superpotential \eqref{potential_Cn_Zn_general}. Notice that these general expressions agree with our convention for the toric phases if we relabel nodes according to $i \to i + 1$ and exchange what we regard as the fundamental fields and their conjugates. Demanding this to hold more generally, results in two additional fusion rules
        \begin{align}
            \mathbf{r}_{ij} \otimes \bar{\mathbf{r}}_{jk} \supseteq \bar{\mathbf{r}}_{ik} & \qquad k < i < j \nonumber\\
            \bar{\mathbf{r}}_{ij} \otimes \mathbf{r}_{jk} \supseteq \bar{\mathbf{r}}_{ik} & \qquad j < k < i   \label{cn_zn_rep_fusion_bar}   
        \end{align} 

Every phase of $\mathbb{C}^{m+2}/\mathbb{Z}_{m+2}$ obeys this structure, i.e. the fields connecting nodes $i$ and $j$ have degree $j-i-1$ and the representations $\mathbf{r}_{ij}$ satisfy the fusion rules \eqref{cn_zn_rep_fusion} and \eqref{cn_zn_rep_fusion_bar}, which determine the superpotential. This can be shown by induction as we now briefly sketch. We will prove this by starting from the toric phase and showing that that if these properties hold for a theory, then they hold for any of its duals. Without loss of generality, we can restrict to a mutation at node $1$.

%=================================================================   
\paragraph{Quiver mutation.} 
%=================================================================    
The incoming chiral fields at node $1$ are in $\Phi^{(0;\mathbf{r}_{01})}_{0,1}$. After the mutation, they become the outgoing chirals
\beq
            \Phi^{(0;\mathbf{r}_{0,1})}_{0,1} \to \Phi^{(0;\bar{\mathbf{r}}_{1,0})}_{1,0} ~.
\eeq 
        Similarly,
\beq
            \Phi^{(i-2;\mathbf{r}_{1,i})}_{1,i} \to \Phi^{(i-1;\mathbf{r}_{1,i})}_{1,i} ~.
\eeq
        Next, we consider the mesons resulting from the composition of $\Phi^{(0;\mathbf{r}_{0,1})}_{0,1}$ and the outgoing arrows at node $1$. They are
\beq
            \Phi^{(0;\mathbf{r}_{0,1})}_{0,1} \circ \Phi^{(i-2;\mathbf{r}_{1,i})}_{1,i} \to \Psi^{(i-2;\mathbf{r}_{0i})}_{0,i} \oplus \Psi^{(i-2;\mathbf{s}_{0,1,i})}_{0,i} ~. \label{cn_zn_mass_terms}
\eeq
        Here we have used the fusion rule \eqref{cn_zn_rep_fusion} to decompose the mesons into two pieces. In the expression above, $\mathbf{s}_{0,1,i}$ is the complement of $\mathbf{r}_{0,i}$ in $\mathbf{r}_{0,1} \otimes \mathbf{r}_{1i}$, i.e.
\beq
            \mathbf{s}_{0,1,i} = (\mathbf{r}_{0,1} \otimes \mathbf{r}_{1,i})/\mathbf{r}_{0,i} ~.
\eeq

%=================================================================   
\paragraph{Superpotential mutation.}
%=================================================================   
The superpotential terms that contain the incoming chiral fields $\Phi^{(0;\mathbf{r}_{0,1})}_{0,1}$ give rise to mass terms of the form
\beq
\Phi_{0,1}^{(0;\mathbf{r}_{0,1})}\Phi_{1,i}^{(i-2;\mathbf{r}_{1,i})}\bar{\Phi}_{i,0}^{(m+1-i;\bar{\mathbf{r}}_{i,0})} \to \Psi^{(i-2;\mathbf{r}_{0i})}_{0,i}\bar{\Phi}_{i,0}^{(m+1-i;\bar{\mathbf{r}}_{i,0})} ~.
\eeq
After integrating out the massive fields, the only surviving fields connected to node $0$ are $\Psi^{(i-2;\mathbf{s}_{01i})}_{0,i}$. As a result, the dual quiver is also monochromatic.

Next, let us consider the terms that involve flavors of node 1 other than $\Phi^{(0;\mathbf{r}_{0,1})}_{0,1}$. They transform as follows
\begin{multline}
                \Phi_{1,1+j}^{(j-2;\mathbf{r}_{1,j+1})}\Phi_{j+1,j+k+1}^{(k-1;\mathbf{r}_{1,k+1})}\bar{\Phi}_{j+k+1,i}^{(m+2-j-k;\bar{\mathbf{r}}_{j+k+1,1})} \\ \to \Phi_{1,1+j}^{(j-1;\mathbf{r}_{1,j+1})}\Phi_{j+1,j+k+1}^{(k-1;\mathbf{r}_{1,k+1})}\bar{\Phi}_{j+k+1,i}^{(m+1-j-k;\bar{\mathbf{r}}_{j+k+1,1})}
\end{multline}
Finally, we add to the superpotential couplings between the mesons and dual flavors
\beq
                \Phi^{(0;\mathbf{r}_{0,1})}_{0,1}\Phi^{(i-1;\mathbf{r}_{1,i})}_{1,i}\bar{\Psi}^{(m-i;\bar{\mathbf{r}}_{i,0})}_{i,0} + \Phi^{(0;\mathbf{r}_{0,1})}_{0,1} \Phi^{(i-1;\mathbf{r}_{1,i})}_{1,i}\bar{\Psi}^{(m-i;\bar{\mathbf{s}}_{i,0,1})}_{i,0} ~.
\label{cn_zn_flavor_meson}
\eeq

In summary, after mutation the new quiver and superpotential are similar in form to the original ones, except for the fact that the dual flavors $\Phi^{(i-1;\mathbf{r}_{1,i})}_{1,i}$ and the surviving mesons $\Psi^{(i-2;\mathbf{s}_{01i})}_{0,i}$ have degrees $(i-1)$ and $(i-2)$, instead of $(i-2)$ and $(i-1)$. This can be accounted for by exchanging the labels of nodes $0$ and $1$. 

In the mirror, this corresponds to moving the vanishing cycle representing node $0$ past the vanishing cycle representing node $1$. After this relabeling, we conclude that under mutation at node $0$ the representations change as
            \begin{align}
                \mathbf{r}_{0,1} &\to \bar{\mathbf{r}}_{0,1} \nonumber\\
                \mathbf{r}_{0,i} &\to \bar{\mathbf{r}}_{1,i} \nonumber\\
                \mathbf{r}_{1,i} &\to \mathbf{s}_{0,1,i} =  (\mathbf{r}_{01} \otimes \mathbf{r}_{1i})/\mathbf{r}_{0i} \nonumber\\
                \mathbf{r}_{i,j} &\to \mathbf{r}_{i,j} && i,j \ne 0,1 \label{mutation_reps_1}
            \end{align} 
It is straightforward to verify that these representations satisfy the fusion rules \eqref{cn_zn_rep_fusion} and \eqref{cn_zn_rep_fusion_bar}.

%=================================================================    
\subsection{Monodromy and Diophantine Equations}
%=================================================================    

In this section we derive the Diophantine equations for the $\mathbb{C}^{m+2}/\mathbb{Z}_{m+2}$ orbifolds. All the dual quivers for these geometries are monochromatic. This implies that the oriented intersection number $S_{i,j}$, which in general is given by the alternating sum in \eref{S_matrix_quiver}, is simply equal, up to a sign, to the number of bifundamental fields between nodes $i$ and $j$. It therefore becomes straightforward to read the quiver from the $S_{i,j}$. We thus get
\beq
S_{i,j} = (-1)^{j-i}\dim(\mathbf{r}_{i,j})~.
\eeq
Using \eqref{mutation_reps_1} and using the dimensions of the representations, we recover the transformation of the $S_{i,j}$ given by \eqref{cycle_crossing}.

The characteristic polynomial $Q(z)$ of the monodromy matrix $M = S^{-T}S$ remains invariant under mutations. We can determine $Q(z)$ for all $\mathbb{C}^{m+2}/\mathbb{Z}_{m+2}$ by focusing on the toric phases discussed in the previous section. Alternatively, we can easily compute it using another result of Cecotti and Vafa. The $2d$ $(2,2)$ theory underlying $\mathbb{C}^{m+2}/\mathbb{Z}_{m+2}$ is the well studied $\mathbb{CP}^{m+1}$ model, whose monodromy matrix has a single Jordan block. As a result, all the eigenvalues of $M$ are equal. As explained in \sref{subsection:dimers_mirror} anomaly cancellation implies that at least one eigenvalue must be $(-1)^{m+1}$. This fixes the characteristic polynomials to be
\beq
Q(z) = \det(z - S^{-T}S) = (z+(-1)^{m})^{m+2} ~.
\label{characteristic_polynomial_orbifold}
\eeq

Below we explicitly present the Diophantine equations for $0 \le m \le 3$. To our knowledge, this is the first time the equations for $m=2$ and $3$ appear in the literature. Obtaining the equations for higher $m$ by expanding \eref{characteristic_polynomial_orbifold} in $z$ is straightforward.

%=================================================================                   
\subsubsection{m=0}
%=================================================================   
 
For $m = 0$, the most general $S$ is
\beq
            S = \begin{pmatrix}
                1 & n_{01} \\
                0 & 1
            \end{pmatrix} ~.
\eeq
The requirement that $Q(z) = (z + 1)^{2}$ implies that
\beq
n_{01} = 2 ~, \label{m_0_monodromy}
\eeq
 which is true for the unique quiver for $\mathbb{C}^{2}/\mathbb{Z}_{2}$. There is no duality for $m=0$. Finally, anomaly cancellation implies that the ranks of the two nodes are equal.

%=================================================================   
\subsubsection{m=1}
%=================================================================   

The equation for $m=1$ was previously studied in \cite{Cachazo:2001sg,Feng:2002kk}. The most general $S$ is\footnote{The minus sign in front of $n_{01}$ and $n_{12}$ corresponds to the $(-1)^{c+1}$ weighting in \eref{S_matrix_quiver}. This convention ensures that the $n_{ij}$ are positive. We will include analogous signs for general $m$. $\mathbb{C}^{m+2}/\mathbb{Z}_{m+2}$ is special in that we obtain the same Diophantine equation regardless of whether we include the $(-1)^{c+1}$ weight or not. This is a consequence of $c+1$ being uniformly $j-i$. For other theories, not including these signs will modify the signs of some of the terms in the resulting Diophantine equations.}
\beq
                S = \begin{pmatrix}
                    1 & -n_{01} & n_{02}  \\
                0 & 1      & -n_{12}  \\
                0 & 0      & 1
                \end{pmatrix}
\eeq
            The requirement that $Q(z)$ is $(z-1)^{3}$ implies the Markov equation 
\beq
                n_{01}^{2} + n_{02}^{2} + n_{12}^{2} - n_{01}n_{02}n_{12} = 0 ~. 
                \label{markov_eq}
\eeq
Every solution to this equation describes a gauge theory associated to $\mathbb{C}^{3}/\mathbb{Z}_{3}$, i.e. a theory connected by a sequence of dualities to the corresponding toric phase discussed in \sref{section_quivers_orbifolds}. The toric phase is the solution to \eref{markov_eq} with
\beq
                n_{01} = n_{02} = n_{12} = 3 ~.
\eeq

%================================================================= 
\paragraph{Ranks.}
%=================================================================   
Anomaly cancellation implies that the ranks are given by integer valued multiples of
\beq
\begin{pmatrix}
                     N_1 \\
                     N_2 \\
                     N_3
                 \end{pmatrix} \propto
                                  \begin{pmatrix}
                     n_{12} \\
                     n_{02} \\
                     n_{01}
                 \end{pmatrix} \label{c3_z3_rank_assignment}
\eeq
This result does not use the Diophantine equation \eref{markov_eq}. In fact, it is possible to show that all the $n_{ij}$'s solving \eref{markov_eq} are multiples of 3. Therefore, as explained in \cite{Cachazo:2001sg,Feng:2002kk}, the most general rank assignment for a given solution takes the form
\beq
\begin{pmatrix}
                     N_1 \\
                     N_2 \\
                     N_3
                 \end{pmatrix} ={N \over 3}
                                  \begin{pmatrix}
                     n_{12} \\
                     n_{02} \\
                     n_{01}
                 \end{pmatrix} \label{c3_z3_rank_assignment}
\eeq
where $N$ is a positive integer and can be regarded as the rank of all nodes in the corresponding toric phase.

%=================================================================   
\subsubsection{m=2}
%=================================================================   

\label{section_ranks_orbifolds_m2}

            For $m=2$ the most general $S$ takes the form
\beq
                S = \begin{pmatrix}
                    1 & -n_{01} & n_{02} & -n_{03} \\
                    0 & 1      & -n_{12} & n_{13} \\
                    0 & 0      & 1 & -n_{23} \\
                    0 & 0 &    0   & 1
                    \end{pmatrix} ~. \label{m2_S}
\eeq
In this case, we get two Diophantine equations from the $z$ expansion of $\det(z-S^{-T}S)=(z+1)^{4}$.
Comparing the coefficients for $z$ (or equivalently $z^{3}$), we get that
  \begin{multline}
                n_{01}^{2} + n_{02}^{2} + n_{03}^{2} + n_{12}^{2} + n_{13}^{2} + n_{23}^{2}
                - n_{01}n_{02}n_{12} \\ - n_{01}n_{03}n_{13} - n_{02}n_{03}n_{23} - n_{12}n_{13}n_{23} 
                + n_{01}n_{03}n_{12}n_{23} = 8 ~,    \label{diophantine_p4_z}
            \end{multline}  
            while comparing the coefficients for $z^{2}$ and using  \eqref{diophantine_p4_z} we get
            \begin{multline}
                n_{01}^2 n_{23}^2 + n_{02}^{2}n_{13}^{2} + n_{03}^{2}n_{12}^{2}    
                 - 2n_{01}n_{02}n_{13}n_{23}  + 2n_{01}n_{03}n_{12}n_{23} - 2n_{02}n_{03}n_{12}n_{13} = 16 ~. \label{diophantine_p4_z2}
            \end{multline}  
            Both sides of this equation are perfect squares. Factorizing it, we can simplify it to
\beq
                (n_{01}n_{23} - n_{02}n_{13} + n_{03}n_{12} + 4)(n_{01}n_{23} - n_{02}n_{13} + n_{03}n_{12} - 4) = 0  ~.
\label{diophantine_p4_z2_factorized}
\eeq
Therefore, the solutions of these Diophantine equations split into two branches, depending on the value of $(n_{01}n_{23} - n_{02}n_{13} + n_{03}n_{12})$. Moreover, it is possible to show that $(n_{01}n_{23} - n_{02}n_{13} + n_{03}n_{12})$ is invariant under mutation, so all the solutions related by duality stay within the same branch. Using the toric phase of $\mathbb{C}^{4}/\mathbb{Z}_{4}$, we deduce that all the theories connected to it by mutations satisfy
\beq
                n_{01}n_{23} - n_{02}n_{13} + n_{03}n_{12} = -4 ~.
\label{diophantine_p4_z2_branch1}
\eeq

As we will explain in detail in \sref{section_C4Z4}, unlike the cases of $m=1$ and $m=0$, not all solutions to \eqref{diophantine_p4_z} and \eqref{diophantine_p4_z2} correspond to $\mathbb{C}^{4}/\mathbb{Z}_{4}$, i.e. there are disconnected mutation webs that satisfy the same equations.\footnote{This phenomenon is not unusual and has been observed for non-orbifold singularities. As we discuss in \sref{section_C4Z4}, the multiplicity of disconnected mutated webs is however far more substantial in this case.}

%================================================================= 
\paragraph{Ranks.}
%=================================================================  

Another important difference with respect to the $m=1$ case are ranks. Following \eref{anomaly_odd}, for $m=1$ the anomaly free ranks are in the null space $S - S^{T}$, which is a $3 \times 3$ antisymmetric matrix. An odd-dimensional antisymmetric matrix always has null vectors, and indeed \eqref{c3_z3_rank_assignment} is a null vector regardless of whether the Markov equation is satisfied. The Markov equation ensures that the monodromy matrix has a single Jordan block and hence that the space of anomaly free ranks is 1-dimensional.

On the other hand, according to \eref{anomaly_even}, for $m=2$ the anomaly free ranks are in the null space of $S + S^{T}$, which is symmetric. Furthermore, in this case it is the Diophantine equations \eref{diophantine_p4_z} and \eref{diophantine_p4_z2} which ensure that anomaly free ranks exist. For theories solving \eref{diophantine_p4_z2_branch1}, the space of anomaly free ranks is 1-dimensional. From \eref{diophantine_p4_z} and \eref{diophantine_p4_z2_branch1}, the ranks are integer multiples of
\beq
\begin{pmatrix}
                     N_1 \\
                     N_2 \\
                     N_3 \\
                     N_4
                 \end{pmatrix} \propto          \begin{pmatrix}
                    2n_{03} + 2n_{12} - n_{01}n_{13} - n_{02}n_{23} + n_{01}n_{12}n_{23} \\
                    2n_{02} - 2n_{13} + n_{01}n_{03} + n_{12}n_{23} \\
                    2n_{01} + 2n_{23} - n_{02}n_{03} - n_{12}n_{13} + n_{01}n_{03}n_{12} \\
                    4 -n_{01}^2 -n_{02}^2 -n_{12}^2 + n_{01}n_{02} n_{12}
                \end{pmatrix} ~.
\eeq
For theories connected to the toric phase of $\mathbb{C}^4/\mathbb{Z}_4$, the proportionality factor is $N/32$, with $N$ a positive integer.

%=================================================================   
\subsubsection{m=3}
%=================================================================    
For $m=3$, the most general $S$ takes the form
\beq
                S = \begin{pmatrix}
                    1 & -n_{01} & n_{02} & -n_{03} & n_{04} \\
                    0 & 1      & -n_{12} & n_{13} & -n_{14} \\
                    0 & 0      & 1 & -n_{23} & n_{24} \\
                    0 & 0 &    0   & 1 & -n_{34} \\
                    0 & 0 &    0   & 0 & 1
                    \end{pmatrix} ~ . \label{m3_S}
\eeq
This time we derive the Diophantine equations from the expansion of $\det(z-S^{-T}S)=(z-1)^{5}$, which gives rise to two independent equations as for $m=2$. The first of them arises from the coefficients of $z$ or $z^{4}$. It is 
\small{
            \begin{align}
                n_{01}^2+n_{02}^2+n_{03}^2+n_{04}^2+n_{13}^2+n_{14}^2+n_{23}^2+n_{24}^2+n_{34}^2 +n_{12}\phantom{abcedefgh}& \nonumber\\
                -n_{01}n_{02}n_{12} - n_{01} n_{03}n_{13} - n_{01} n_{04}n_{14} - n_{02} n_{03}n_{23} -n_{02} n_{04}n_{24}& \nonumber\\
                 - n_{03} n_{04}n_{34} -n_{12}n_{13}n_{23} - n_{12}n_{14} n_{24} - n_{13} n_{14} n_{34}-n_{23} n_{24} n_{34}& \nonumber\\
                +n_{01} n_{03} n_{12}n_{23} + n_{01} n_{04}n_{12} n_{24} + n_{01} n_{04}n_{13} n_{34}&  \nonumber\\ 
                + n_{01} n_{04}n_{13} n_{34} + n_{02} n_{04} n_{23} n_{34}  +n_{12}n_{14} n_{23} n_{34}& \nonumber\\
                -n_{01}n_{12}n_{23} n_{34}n_{04} &= 0 ~.
            \end{align}}
            The second equation, which comes from the coefficients of $z^{2}$ or $z^{3}$, is
\small{
            \begin{align}
                n_{34}^2 n_{12}^2+ n_{03}^2 n_{12}^2+n_{04}^2 n_{12}^2 +n_{14}^2 n_{23}^2+n_{13}^2 n_{24}^2+n_{23}^2 n_{01}^2+n_{24}^2 n_{01}^2+n_{34}^2 n_{01}^2\phantom{abcdefghijklmno}& \nonumber\\
                +n_{14}^2 n_{03}^2+n_{24}^2 n_{03}^2+n_{13}^2 n_{04}^2+n_{23}^2 n_{04}^2 + n_{13}^2 n_{02}^2+n_{14}^2 n_{02}^2+n_{34}^2 n_{02}^2 \phantom{abcedfghijklmno}&\nonumber \\
                +2 n_{14} n_{23} n_{34} n_{12}-2 n_{13} n_{24} n_{34} n_{12}+2 n_{23} n_{01} n_{03} n_{12}-2 n_{13} n_{02} n_{03} n_{12}+2 n_{24} n_{01} n_{04} n_{12}&\nonumber\\
                -2 n_{14} n_{02} n_{04} n_{12}-2 n_{13} n_{14} n_{23} n_{24}-2 n_{13} n_{23} n_{01} n_{02}-2 n_{14} n_{24} n_{01} n_{02}-2 n_{14} n_{34} n_{01} n_{03}&\nonumber\\
                -2 n_{24} n_{34} n_{02} n_{03}+2 n_{13} n_{34} n_{01} n_{04}+2 n_{23} n_{34} n_{02} n_{04}-2 n_{13} n_{14} n_{03} n_{04}-2 n_{23} n_{24} n_{03} n_{04}&\nonumber\\
                -n_{34} n_{03} n_{04} n_{12}^2-n_{14} n_{24} n_{03}^2 n_{12}-n_{13} n_{23} n_{04}^2 n_{12} -n_{23} n_{24} n_{34} n_{01}^2-n_{13} n_{14} n_{34} n_{02}^2 &\nonumber\\
                -n_{13} n_{24}^2 n_{01} n_{03} - n_{14} n_{23}^2 n_{01} n_{04}-n_{13}^2 n_{24} n_{02} n_{04} -n_{14}^2 n_{23} n_{02} n_{03}-n_{34}^2 n_{01} n_{02} n_{12} &\nonumber\\
                +n_{24} n_{34} n_{01} n_{03} n_{12}+n_{14} n_{34} n_{02} n_{03} n_{12}+n_{13} n_{24} n_{03} n_{04} n_{12} +n_{13} n_{34} n_{02} n_{04} n_{12}&\nonumber\\
                + n_{14} n_{23} n_{03} n_{04} n_{12}+n_{14} n_{23} n_{34} n_{01} n_{02}+n_{13} n_{24} n_{34} n_{01} n_{02}+n_{14} n_{23} n_{24} n_{01} n_{03}& \nonumber\\
                +n_{13} n_{14} n_{24} n_{02} n_{03}+n_{13} n_{23} n_{24} n_{01} n_{04} +n_{13} n_{14} n_{23} n_{02} n_{04} -3 n_{23} n_{34} n_{01} n_{04} n_{12} =0
            \end{align}}

%=================================================================   
\paragraph{Anomaly Free Ranks}
%=================================================================   

As for $m=1$, anomaly free ranks are in the null space if $S - S^{-T}$ and can hence be determined without any reference to the above Diophantine equations. They are integer multiples of
\beq
            \begin{pmatrix}
                     N_1 \\
                     N_2 \\
                     N_3 \\
                     N_4 \\
                     N_5
                 \end{pmatrix} \propto     
                \begin{pmatrix}
                -n_{14} n_{23} + n_{13} n_{24} - n_{12} n_{34} \\
                -n_{04} n_{23} + n_{03} n_{24} - n_{02} n_{34} \\ 
                -n_{04} n_{13} + n_{03} n_{14} - n_{01} n_{34} \\
                -n_{04} n_{12} + n_{02} n_{14} - n_{01} n_{24} \\ 
                -n_{03} n_{12} + n_{02} n_{13} - n_{01} n_{23}
                \end{pmatrix} ~.
\eeq            
For the toric phase of $\mathbb{C}^5/\mathbb{Z}_5$, the proportionality constant is $N/25$, with $N$ a positive integer.

%=================================================================
\section{Classification of Solutions - A Case Study: $\mathbb{C}^4/\mathbb{Z}_4$}
%=================================================================

\label{section_C4Z4}

In this section we consider the classification of solutions to the Diophantine equations, focusing on the $\mathbb{C}^4/\mathbb{Z}_4$ example to illustrate our ideas.

%=================================================================
\subsection{Ordering the Nodes}
%=================================================================

\label{section_ordering_nodes}

The discussion in the previous section requires the nodes of the quiver to be ordered according to the degree of the fields connecting them. This ordering coincides with the one arising from the mirror geometry, which we explained in \sref{section_mirror_grading}. Since a mutation corresponds to a reorganization of the vanishing cycles in the mirror, after dualization it is necessary to relabel the nodes of the quiver accordingly. 

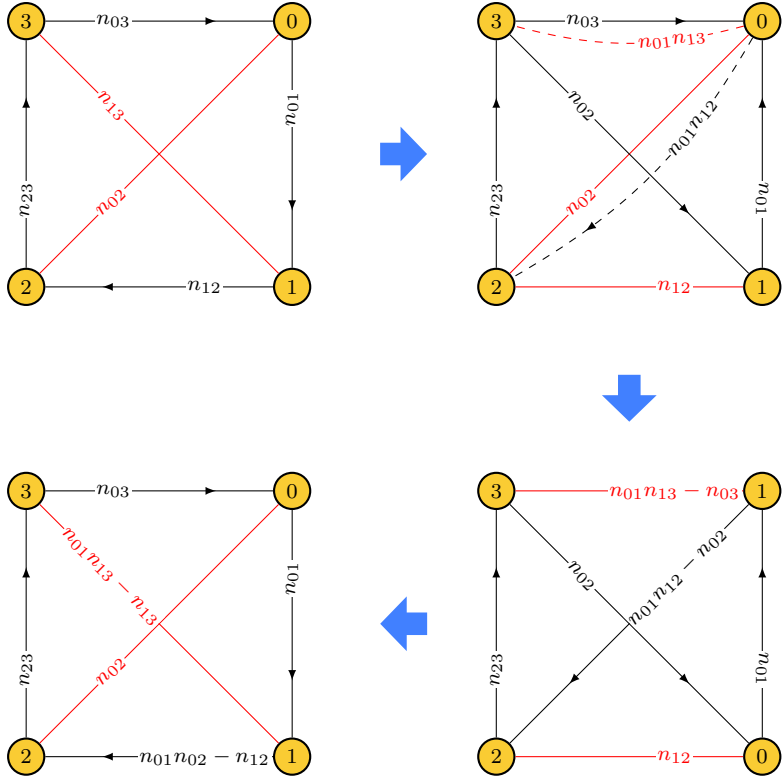
\begin{figure}[ht]
        \centering
         \newcommand{\pos}{0.3}
            \newcommand{\arrowHeadPosition}{0.75}
            \begin{tikzpicture}[scale=2.5 , decoration={markings,mark=at position \arrowHeadPosition with {\arrow{latex}}}] 
                \draw[draw=white] (0,0) circle (1);
                \tikzstyle{every node}=[circle,thick,fill=yellow2,draw,inner sep=2.5pt,font=\scriptsize]
                    \draw(0.7071067811865474,0.7071067811865477) node ("A") {$0$};
                    \draw(0.7071067811865477,-0.7071067811865474) node ("B") {$1$};
                    \draw(-0.7071067811865467,-0.7071067811865483) node ("C") {$2$};
                    \draw(-0.7071067811865485,0.7071067811865466) node ("D") {$3$};
                    \draw [postaction={decorate}, black]("A")tonode[pos = \pos, draw = none , fill = white , inner sep = 0.5, rectangle,sloped]{$n_{01}$}("B");
                    \draw [red]("C")to node[pos = \pos, draw = none , fill = white , inner sep = 0.5, rectangle,sloped]{$n_{02}$}("A");
                    \draw [postaction={decorate}, black]("D")tonode[pos = \pos, draw = none , fill = white , inner sep = 0.5, rectangle,sloped]{$n_{03}$}("A");
                    \draw [postaction={decorate}, black]("B")tonode[pos = \pos, draw = none , fill = white , inner sep = 0.5, rectangle,sloped]{$n_{12}$}("C");
                    \draw [red]("D")to node[pos = \pos, draw = none , fill = white , inner sep = 0.5, rectangle, sloped]{$n_{13}$}("B");
                    \draw [postaction={decorate}, black]("C")tonode[pos = \pos, draw = none , fill = white , inner sep = 0.5, rectangle,sloped]{$n_{23}$}("D");
                    \draw [-{Triangle[angle = 90:5mm]},line width = 3mm,blue2](1.175,0) -- (1.425,0);
                    \begin{scope}[shift={(2.5,0)}]
                        \draw(0.7071067811865474,0.7071067811865477) node ("A") {$0$};
                        \draw(0.7071067811865477,-0.7071067811865474) node ("B") {$1$};
                        \draw(-0.7071067811865467,-0.7071067811865483) node ("C") {$2$};
                        \draw(-0.7071067811865485,0.7071067811865466) node ("D") {$3$};
                        \draw [postaction={decorate}, black]("B")tonode[pos = \pos, draw = none , fill = white , inner sep = 0.5, rectangle,sloped]{$n_{01}$}("A");
                        \draw [red]("C")to node[pos = \pos, draw = none , fill = white , inner sep = 0.5, rectangle,sloped]{$n_{02}$}("A");
                        \draw [postaction={decorate}, black]("D")tonode[pos = \pos, draw = none , fill = white , inner sep = 0.5, rectangle,sloped]{$n_{03}$}("A");
                        \draw [red]("B")tonode[pos = \pos, draw = none , fill = white , inner sep = 0.5, rectangle,sloped]{$n_{12}$}("C");
                        \draw [postaction = {decorate} , black]("D")to node[pos = \pos, draw = none , fill = white , inner sep = 0.5, rectangle, sloped]{$n_{02}$}("B");
                        \draw [postaction={decorate}, black]("C")tonode[pos = \pos, draw = none , fill = white , inner sep = 0.5, rectangle,sloped]{$n_{23}$}("D");
                        \draw [postaction={decorate}, black, bend left = 15 , dashed]("A")tonode[pos = \pos, draw = none , fill = white , inner sep = 0.5, rectangle,sloped ]{$n_{01}n_{12}$}("C");
                        \draw [red,dashed,bend left = 15]("A")to node[pos = \pos, draw = none , fill = white , inner sep = 0.5, rectangle,sloped]{$n_{01}n_{13}$}("D");
                    \end{scope}
                    \draw [-{Triangle[angle = 90:5mm]},line width = 3mm,blue2](2.5,-1.175) -- (2.5,-1.425);
                    \begin{scope}[shift={(2.5,-2.5)}]
                        \draw(0.7071067811865474,0.7071067811865477) node ("A") {$1$};
                        \draw(0.7071067811865477,-0.7071067811865474) node ("B") {$0$};
                        \draw(-0.7071067811865467,-0.7071067811865483) node ("C") {$2$};
                        \draw(-0.7071067811865485,0.7071067811865466) node ("D") {$3$};
                        \draw [postaction={decorate}, black]("B")tonode[pos = \pos, draw = none , fill = white , inner sep = 0.5, rectangle,sloped]{$n_{01}$}("A");
                        \draw [red]("B")tonode[pos = \pos, draw = none , fill = white , inner sep = 0.5, rectangle,sloped]{$n_{12}$}("C");
                        \draw [postaction = {decorate} , black]("D")to node[pos = \pos, draw = none , fill = white , inner sep = 0.5, rectangle, sloped]{$n_{02}$}("B");
                        \draw [postaction={decorate}, black]("C")tonode[pos = \pos, draw = none , fill = white , inner sep = 0.5, rectangle,sloped]{$n_{23}$}("D");
                        \draw [postaction={decorate}, black]("A")tonode[pos = \pos, draw = none , fill = white , inner sep = 0.5, rectangle,sloped ]{$n_{01}n_{12}-n_{02}$}("C");
                        \draw [red]("A")to node[pos = \pos, draw = none , fill = white , inner sep = 0.5, rectangle,sloped]{$n_{01}n_{13}-n_{03}$}("D");
                    \end{scope}
                    \draw [-{Triangle[angle = 90:5mm]},line width = 3mm,blue2](1.425,-2.5) -- (1.175,-2.5);
                    \begin{scope}[shift={(0,-2.5)}]
                        \draw(0.7071067811865474,0.7071067811865477) node ("A") {$0$};
                        \draw(0.7071067811865477,-0.7071067811865474) node ("B") {$1$};
                        \draw(-0.7071067811865467,-0.7071067811865483) node ("C") {$2$};
                        \draw(-0.7071067811865485,0.7071067811865466) node ("D") {$3$};
                        \draw [postaction={decorate}, black]("A")tonode[pos = \pos, draw = none , fill = white , inner sep = 0.5, rectangle,sloped]{$n_{01}$}("B");
                        \draw [red]("C")to node[pos = \pos, draw = none , fill = white , inner sep = 0.5, rectangle,sloped]{$n_{02}$}("A");
                        \draw [postaction={decorate}, black]("D")tonode[pos = \pos, draw = none , fill = white , inner sep = 0.5, rectangle,sloped]{$n_{03}$}("A");
                        \draw [postaction={decorate}, black]("B")tonode[pos = \pos, draw = none , fill = white , inner sep = 0.5, rectangle,sloped]{$n_{01}n_{02} - n_{12}$}("C");
                        \draw [red]("D")to node[pos = \pos, draw = none , fill = white , inner sep = 0.5, rectangle, sloped]{$n_{01}n_{13}-n_{13}$}("B");
                        \draw [postaction={decorate}, black]("C")tonode[pos = \pos, draw = none , fill = white , inner sep = 0.5, rectangle,sloped]{$n_{23}$}("D");
                    \end{scope}
            \end{tikzpicture} 
            \caption{A mutation for $\mathbb{C}^{4}/\mathbb{Z}_{4}$. The first figure is the original quiver. Mutating on node $1$ we obtain the second quiver, where dashed arrows represent mesons. In the third step we integrated out massive fields and exchanged the labels of nodes $0$ and $1$. In the last step we switched the positions of nodes $0$ and $1$.}
            \label{c4z4_mutation_node_1}
    \end{figure}

We saw that for $\mathbb{C}^{4}/\mathbb{Z}_{4}$ the nodes in the quiver can be ordered such that the degree of arrows connecting nodes $i$ and $j$ is $j-i-1$ if $j > i$. \fref{c4z4_mutation_node_1} illustrates the corresponding relabeling in an explicit example. It shows how, after a mutation on node $1$, the appropriate order is restored by switching the labels of nodes $0$ and $1$. Similarly, after a mutation on node $i$ we need to exchange the labels of nodes $i-1$ and $i$. It is only after this reordering that the monodromy equations \eref{diophantine_p4_z} and \eref{diophantine_p4_z2} are satisfied.

%=================================================================
%=================================================================

%=================================================================
\subsection{Classification of Seeds}
%=================================================================

\label{section_classification_seeds}

In general, the Diophantine equations associated to a given singularity have an infinite number of solutions. Therefore, their classification may seem to be a monumental problem. An ingenious approach to partially addressing this question was introduced in \cite{Cecotti:1992rm}, which proposed to focus on the {\it seeds} of duality webs. A seed is defined as a ``minimal" theory within a mutation class, where minimality can be loosely defined as having the ``smallest size", as measured by the ranks of gauge groups and the matter content. In practice, we are simply interested in finding small representatives of every mutation class.

In the context of toric singularities, it is natural to search for seeds that satisfy anomaly cancellation conditions with equal ranks for all gauge groups.\footnote{It is worth noting that the notion of {\it toric phase} is more restrictive than all gauge groups having the same rank. A toric phase is a quiver theory that is associated to an $m$-dimer \cite{Franco:2019bmx}. In particular, some of the seeds we obtain by requiring equal ranks may not be described by $m$-dimers.} Generically, there are multiple such theories for a given toric CY $(m+2)$-fold. Therefore, in order to identify a pair of them as independent seeds, it is necessary to verify that they are not connected by a sequence of mutations. 

To illustrate these ideas, let us classify the equal rank seeds for the Diophantine equations for $\mathbb{C}^4/\mathbb{Z}_4$, \eref{diophantine_p4_z} and \eref{diophantine_p4_z2}. In \sref{section_independence_seeds} we will show that each of these theories indeed generates an independent mutation web and therefore is a seed.\footnote{A priori, it is logically possible that non-equal rank seeds exist. In such a case, there would be additional solutions, belonging to disconnected duality webs that do not contain any equal rank quiver. We will not explore this possibility.}

The most general intersection matrix $S$ that satisfies anomaly cancellation condition with equal ranks can be parameterized as:
\beq
            S = \begin{pmatrix}
                    1 & -\tilde{n} & \tilde{n}+n-2 & -n \\
                    0 & 1 & -n & \tilde{n}+n-2  \\
                    0 & 0 & 1 & -\tilde{n} \\
                    \ 0 \ \ & 0 & 0 & 1
                \end{pmatrix}   ~.
\eeq                
Expressing \eref{diophantine_p4_z} and \eref{diophantine_p4_z2} in terms of $n$ and $\tilde{n}$ leads to a remarkable simplification. Both of them give rise to the same equation, which takes the form
\beq
(n-2)(\tilde{n}-2)(n\tilde{n}-2n-2\tilde{n}) = 0 ~. 
\label{reduced_diophantne}   
\eeq
Notice that exchanging $n$ and $\tilde{n}$ results in the same quiver up to a cyclic reordering of nodes. We will therefore consider the solutions related by this exchange as equivalent. With this in mind the distinct solutions to this Diophantine equation with positive $\tilde{n}$ and $n$ are as follows.

\medskip

%=================================================================
\paragraph{Two isolated seeds.} 
%=================================================================

There are two isolated solutions, which correspond to the positive integer roots of the second factor in \eref{reduced_diophantne}, i.e. $(n\tilde{n}-2n-2\tilde{n})$. They are:
\begin{enumerate}
                    \item[{\bf 1.}]
                        \underline{$\tilde{n} = n = 4$}. The corresponding $S$ becomes
                        \beq
                            S_{t} = \begin{pmatrix}
                                1 & -4 & 6 & -4 \\
                                0 &  1 & -4 & 6 \\
                                0 & 0  & 1 & -4 \\
                                0 & 0 & 0 & 1
                            \end{pmatrix} ~.
                        \eeq
                        This solution corresponds to the toric phase of $\mathbb{C}^{4}/\mathbb{Z}_{4}$, therefore we indicate it with a subscript $t$.
                        
                    \item[{\bf 2.}] \underline{$\tilde{n}=3$ and $n = 6$}. For this solution, we have
			\beq
                            S_{e} = \begin{pmatrix}
                                        1 & -3 & 7 & -6 \\
                                        0 & 1 & -6 & 7 \\
                                        0 & 0 & 1 & -3 \\
                                        0 & 0 & 0 & 1 
                                     \end{pmatrix} ~.
                        \eeq
                        The subscript $e$ denotes `exceptional' since this is the only solution, other than $S_t$, that does not belong to the infinite family that we discuss below.
                \end{enumerate}

\medskip

%=================================================================
\paragraph{An infinite family of seeds.}
%=================================================================

                This family corresponds to $\tilde{n} = 2$ and $n$ an arbitrary positive integer. This results in an infinite family $S_{n}$ of seeds of the form
                \beq
                    S_{n} = \begin{pmatrix}
                                1 & -2 & n & -n \\
                                0 & 1 & -n & n \\
                                0 & 0 & 1 & -2 \\
                                0 & 0 & 0 & 1
                            \end{pmatrix} ~.
                \eeq
For $n=1$, there are a pair of nodes with a single incoming chiral field. If all ranks are equal, naive application of triality to any of these two nodes would result in their disappearance. As mentioned in \sref{section_additional_conditions_mutations}, this situation is the $2d$ analogue of an $N_f=N_c$ node in $4d$ $\mathcal{N}=1$ gauge theories. We will thus restrict to $n\geq 2$.

%=================================================================
\subsection{Independence of the Seeds}
%=================================================================

\label{section_independence_seeds}

A priori, it is possible that our previous analysis misinterpreted some of the seeds, namely that they are not indeed independent theories but are instead connected by sequences of mutations. We now show that this is not the case.

We first recall that factorization of \eref{diophantine_p4_z2} given by \eref{diophantine_p4_z2_branch1}. Since $(n_{01} n_{23} - n_{02}n_{13} + n_{03}n_{12})$ is mutation invariant, it splits the space of solutions into two disconnected branches. Evaluating it for the seeds described above we get that 
\begin{align}
                 S_n&: n_{01} n_{23} - n_{02}n_{13} + n_{03}n_{12} = 4 \nonumber \\
                 S_{t},S_{e}&: n_{01} n_{23} - n_{02}n_{13} + n_{03}n_{12} = -4
\end{align}
Therefore, $S_t$ and $S_e$ are on one branch, while the infinity family $S_n$ lies on the other branch.

%=================================================================
\paragraph{$S_{t}$ and $S_{e}$ are disconnected.} 
%=================================================================
We note that $S_{e}$ has both even and odd off-diagonal entries and the same is true for all solutions related to it by mutations. On the other hand, $S_{t}$ and all the solutions obtained from it by mutations only have even off-diagonal entries. The preservation of these properties under mutation simply follows from the transformation of flavors and how composite mesons are generated. We thus conclude that $S_{t}$ and $S_{e}$ are indeed independent seeds.

%=================================================================
\paragraph{$S_n$ and $S_{n'}$ are disconnected.} 
%=================================================================
Now we turn to the question of whether two members of the infinite family, $S_n$ and $S_{n'}$ with $n\neq n'$, can be connected by a sequence of mutations. 

First of all, $S_2$ is self-dual under mutation on any of its nodes, hence it generates a ``duality web" consisting of a single element and it is disconnected from other members of the $S_n$ family. 

To make further progress, we generalize the argument we used to show that $S_{t}$ and $S_{e}$ are disconnected. There we exploited the fact that if the $n_{ij}$ satisfy the Diophantine equations, they also satisfy them modulo 2 or, more generally, modulo any integer $k$. When the $n_{ij}$ are considered as integers, the duality web can be infinite and two elements in it can be related by an arbitrarily long sequence of mutations. On the other hand, the duality web becomes finite when the $n_{ij}$ are regarded as elements of $\mathbb{Z}_{k}$. Furthermore, we can optimize it to be small by appropriately choosing $k$. If two seeds are related by a sequence of mutations, then the $\mathbb{Z}_{k}$ mutation webs generated by them must be the same for any integer $k$. This argument allows us to distinguish between the web generated by $S_n$ and $S_{n'}$. The $\mathbb{Z}_{n-2}$ web generated by $S_{n}$ has a single element, i.e.
\beq
                 \begin{pmatrix}
                     1 & -2 & 2 & -2 \\
                     0 & 1 & -2 & 2 \\
                     0 & 0 & 1 & -2 \\
                     0 & 0 & 0 & 1
                 \end{pmatrix} \mod (n-2) ~.
\eeq
Therefore, if $S_n$ and $S_{n'}$ are related by a sequence of mutations we must have that $n'-2$ divides $n-2$. Exchanging the roles of $n$ and $n'$ and running the same argument implies that $n-2$ divides $n'-2$. Hence, $n=n'$.
This completes the proof that if $n\neq n'$, $S_n$ and $S_{n'}$ are independent seeds and they generate disconnected duality webs.

%=================================================================
\paragraph{Further thoughts on infinite seeds.} 
%=================================================================

Let us briefly reflect on what we have just shown: the Diophantine equations associated to the $\mathbb{C}^4/\mathbb{Z}_4$ geometry admit an infinite number of seeds. As expected, one of them is the toric phase for $\mathbb{C}^4/\mathbb{Z}_4$. The multiplicity of seeds is not surprising. In fact, there are even explicit known examples of theories realized on D-branes probing different CY singularities that are associated to the same sets of Diophantine equations. Furthermore, examples of Diophantine equations with an infinite number of seeds were already found in \cite{Cecotti:1992rm}. The physical interpretation of all these theories is, however, unknown.

The infinite class of seeds we have found is novel in a variety of ways. First of all, to our knowledge, this is the first family of seeds that are explicitly $m=2$ graded quivers, i.e. $2d$ $\mathcal{N}=(0,2)$ theories. Moreover, in contrast with the infinite family in \cite{Cecotti:1992rm}, all these theories satisfy anomaly cancellation with equal ranks. 

In \sref{section_seeds_anomalies} and \sref{section_seeds_webs} we will continue with the characterization of these seeds, identifying properties that distinguish between them, and in \sref{section_seeds_interpretation} we will investigate the physical realization of some of these theories.

%=================================================================
\subsection{Anomaly Free Rank Assignments}
%=================================================================

\label{section_seeds_anomalies}

Let us now consider the anomaly free rank assignments for the seeds we have found. We will see that the dimensionality of the space of anomaly free rank assignments distinguishes between the isolated and the infinite families of seeds.

%=================================================================
\paragraph{$S_t$ and $S_e$.}
%=================================================================

We partially dealt with this question in \sref{section_ranks_orbifolds_m2} where we used \eqref{diophantine_p4_z} and \eqref{diophantine_p4_z2_branch1} to find the anomaly free ranks for the entire mutation web generated by $S_{t}$. Since any dual of $S_{e}$ also satisfies both of these equations, the anomaly free ranks for all the theories in its mutation web are given by the same functions of the $n_{ij}$. In particular, this means that for the seeds $S_{t}$ and $S_{e}$, the anomaly free rank assignments are integer multiples of:
\be
(1,1,1,1) \, ,
\label{regular_brane_ranks}
\eeq
where, for convenience, we have switched to represent the ranks as row vectors.

%=================================================================
\paragraph{$S_n$ family.}
%=================================================================

In contrast, for $S_n$ the space of allowed ranks is no longer 1-dimensional.  For $n> 2$, the space of anomaly free ranks is spanned by
\beq
\begin{array}{c}
(1,1,1,1) \\[.1 cm]
(0,0,1,1)
\end{array}
\eeq            
The case of $n=2$ is special, since the space of anomaly free ranks is $3$-dimensional. It is spanned by
\beq
\begin{array}{c}
(1,1,1,1) \\[.1 cm]
(0,0,1,1) \\[.1 cm]
(1,0,0,1) 
\end{array}
\eeq

%=================================================================
\subsection{Further Analysis: Interpretation of the Other Seeds}
%=================================================================

\label{section_seeds_interpretation}

In this section we investigate the possible physical interpretation of the infinite set of seeds we obtained for the Diophantine equations associated to the $\mathbb{C}^4/\mathbb{Z}_4$ orbifold. Rather than looking for the most general possible realization of these theories, we will restrict to the very concrete corner of theories that can be engineered on D1-branes probing toric CY 4-folds. For this class of geometries, the problem is well defined and powerful tools for addressing it exist.

Since the quivers under consideration have four gauge groups, we should focus on toric diagrams, which for CY 4-folds are $3d$ convex lattice polytopes, with normalized volume equal to 4.\footnote{The normalization is with respect to the volume of a minimal $3d$ integer tetrahedron.}  It was first proved in \cite{MR1138580} that, up to unimodular equivalence, there are finitely many $d$-dimensional convex lattice polytopes having volume lower than a constant $K$. Then, \cite{balletti2018enumeration} introduced an algorithm for the complete enumeration of such equivalence classes for arbitrary $d$ and $K$. The author produced a large number of classes of polytopes with this algorithm, which are available at \cite{linkpolytopes}.

There are 17 lattice polytopes of volume 4, which are given by:
\beq
\begin{array}{rl}
{\bf 1.} & (0,0,0) \, , \, (1,0,0) \, , \, (0,1,0) \, , \, (3,3,4)  \\[.05cm]
{\bf 2.} & (0,0,0) \, , \, (1,0,0) \, , \, (0,1,0) \, , \, (0,1,4)  \\[.05cm]
{\bf 3.} & (0,0,0) \, , \, (1,0,0) \, , \, (1,2,0) \, , \, (1,0,2)  \\[.05cm]
{\bf 4.} & (0,0,0) \, , \, (1,0,0) \, , \, (0,1,0) \, , \, (2,1,4)  \\[.05cm]
{\bf 5.} & (0,0,0) \, , \, (1,0,0) \, , \, (0,1,0) \, , \, (1,1,4)  \\[.05cm]
{\bf 6.} & (0,0,0) \, , \, (1,0,0) \, , \, (0,2,0) \, , \, (1,0,2)  \\[.05cm]
{\bf 7.} & (0,0,0) \, , \, (1,0,0) \, , \, (0,1,0) \, , \, (2,3,4)  \\[.05cm]
{\bf 8.} & (0,0,0) \, , \, (1,0,0) \, , \, (0,1,0) \, , \, (0,0,1) \, , \, (2,2,-1)  \\[.05cm]
{\bf 9.} & (0,0,0) \, , \, (1,0,0) \, , \, (0,1,0) \, , \, (1,1,2) \, , \, (1,2,2)  \\[.05cm]
{\bf 10.} & (0,0,0) \, , \, (1,0,0) \, , \, (0,1,0) \, , \, (0,0,1) \, , \, (1,-2,-1)  \\[.05cm]
{\bf 11.} & (0,0,0) \, , \, (1,0,0) \, , \, (1,2,0) \, , \, (0,0,1) \, , \, (2,2,-1)  \\[.05cm]
{\bf 12.} & (0,0,0) \, , \, (1,0,0) \, , \, (0,1,0) \, , \, (0,0,1) \, , \, (3,-3,1)  \\[.05cm]
{\bf 13.} & (0,0,0) \, , \, (1,0,0) \, , \, (1,2,0) \, , \, (0,0,1) \, , \, (2,1,-1)  \\[.05cm]
{\bf 14.} & (0,0,0) \, , \, (1,0,0) \, , \, (0,1,0) \, , \, (1,0,2) \, , \, (0,1,-2)  \\[.05cm]
{\bf 15.} & (0,0,0) \, , \, (1,0,0) \, , \, (0,1,0) \, , \, (-1,1,0) \, , \, (0,0,1) \, , \, (-1,2,-1) \\[.05cm]
{\bf 16.} & (0,0,0) \, , \, (1,0,0) \, , \, (0,1,0) \, , \, (1,1,0) \, , \, (0,0,1) \, , \, (0,2,1) \\[.05cm]
{\bf 17.} & (0,0,0) \, , \, (1,0,0) \, , \, (0,1,0) \, , \, (0,0,1) \, , \, (-1,1,1) \, , \, (0,1,1)
\end{array}
\label{toric_diagrams_volume_4}
\eeq
where we only present the corners of the toric diagrams. Since there are only 17 toric diagrams in this list, we already conclude that not all the infinite seeds we found can be realized in this context.

The first toric diagram is $SL(3,\mathbb{Z})$ equivalent to 
\beq
(1,0,0) \, , \, (0,1,0) \, , \, (0,0,1) \, , \, (-1,-1,-1) \, ,
\eeq
so it corresponds to the $\mathbb{C}^4/\mathbb{Z}_4$ orbifold under consideration.

Remarkably, determining which of the toric diagrams in \eref{toric_diagrams_volume_4} correspond to some of the seeds we have found is a tractable problem, thanks to the efficient tools developed in recent years for constructing $2d$ $\mathcal{N}=(0,2)$ gauge theories on D1-branes probing singularities. These methods include partial resolution, orbifold reduction and $3d$ printing and we refer reader to \cite{Franco:2015tna,Franco:2016fxm,Franco:2018qsc} for details. Our strategy is as follows:
\begin{itemize}
\item We will construct one toric phase for each of the 17 toric diagrams in \eref{toric_diagrams_volume_4} using the aforementioned procedures.
\item For each of these theories, we will check whether they satisfy the Diophantine equations \eref{diophantine_p4_z} and \eref{diophantine_p4_z2}.
\item If they do, they must correspond to one of our seeds.
\end{itemize}
Note that we have only specified the quivers but not the superpotentials for the seeds. Even when restricting to superpotentials satisfying the toric condition, there might be multiple possible superpotentials, and hence toric CY$_4$'s, consistent with the same quiver.\footnote{The toric condition is a restriction on the structure of the superpotential of a theory associated to a toric CY $(m+2)$-fold. See \cite{Franco:2015tna} for a discussion of $m=2$ and \cite{Franco:2019bmx} for general $m$.}

%=================================================================
\subsection*{Results}
%=================================================================

After constructing gauge theories for the 17 toric CY 4-folds in \eref{toric_diagrams_volume_4}, it turns out that 5 of them correspond to seeds in our classification, as we explain below. For the remaining 12 toric diagrams, all toric phases contain adjoint fields, so they do not fit into our classification.

\begin{figure}[ht]
	\centering
	\includegraphics[width=4.5cm]{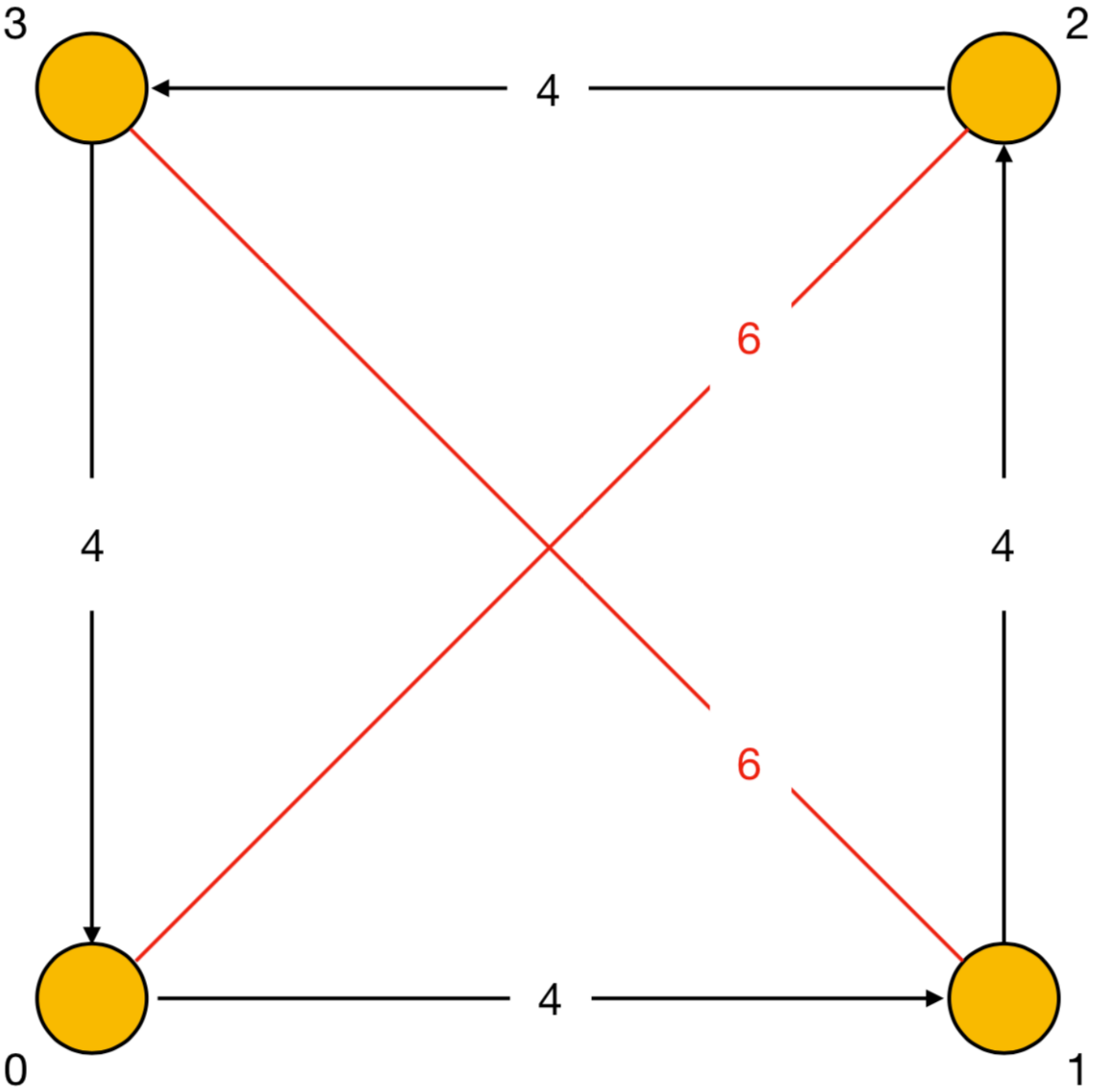}
\caption{Quiver diagram for $\mathbb{C}^4/\mathbb{Z}_4$.}
	\label{1_c4z4}
\end{figure}

%=================================================================
\paragraph{\underline{$S_t$}.}
%=================================================================
As already mentioned, the first toric diagram in \eref{toric_diagrams_volume_4} corresponds to $\mathbb{C}^4/\mathbb{Z}_4$, our starting point. The gauge theory for this orbifold is well known (see e.g. \cite{Franco:2015tna}), its quiver is shown in \fref{1_c4z4} and it obviously corresponds to $S_t$.

%=================================================================
%=================================================================

\begin{figure}[h]
\begin{center}
\begin{subfigure}{0.4\textwidth}
  \centering
  \includegraphics[width=.7\linewidth]{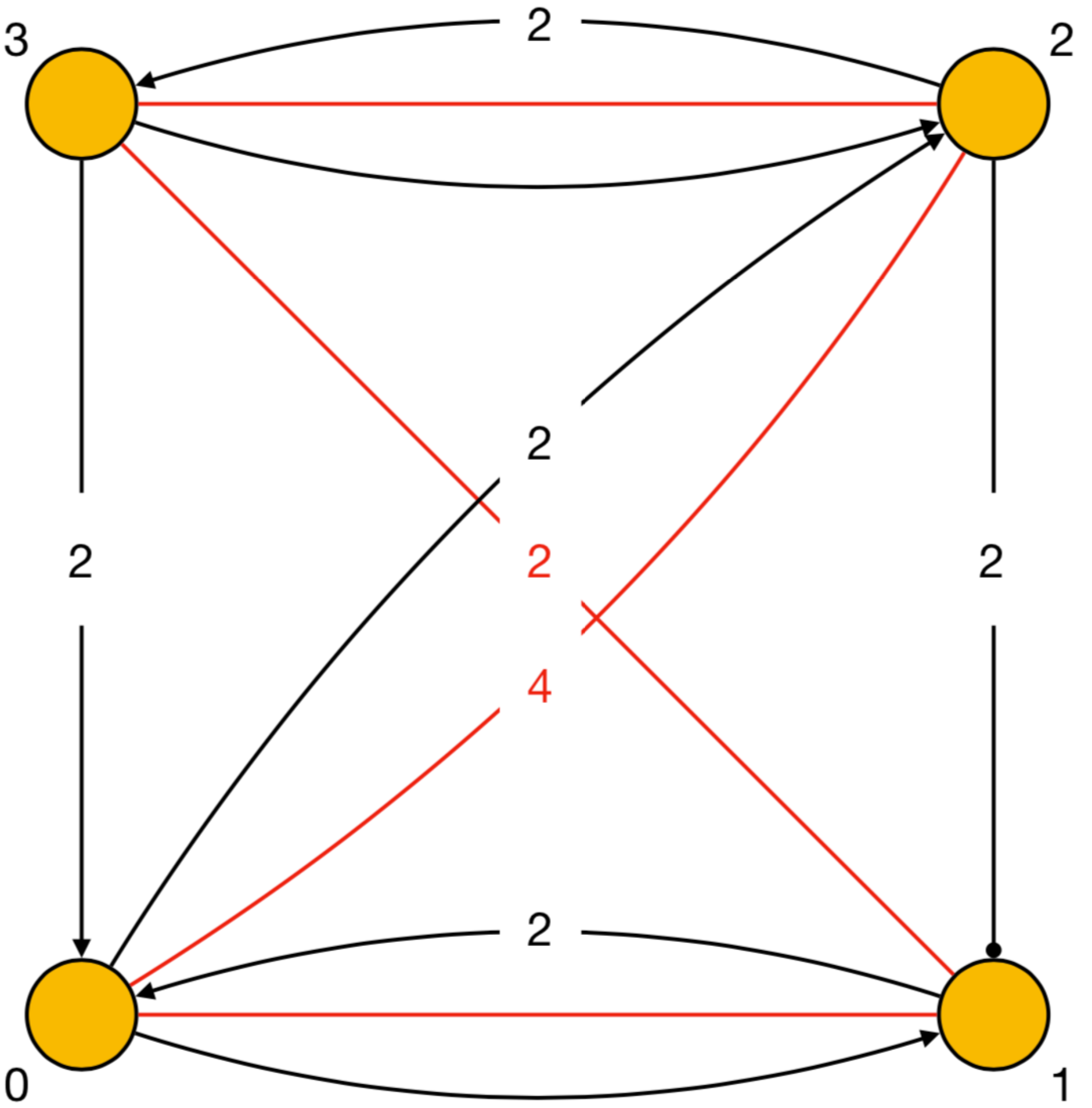}\quad
  \caption{Model 8}
  \label{fig:1}
\end{subfigure}%
\begin{subfigure}{0.4\textwidth}
  \centering
  \includegraphics[width=.7\linewidth]{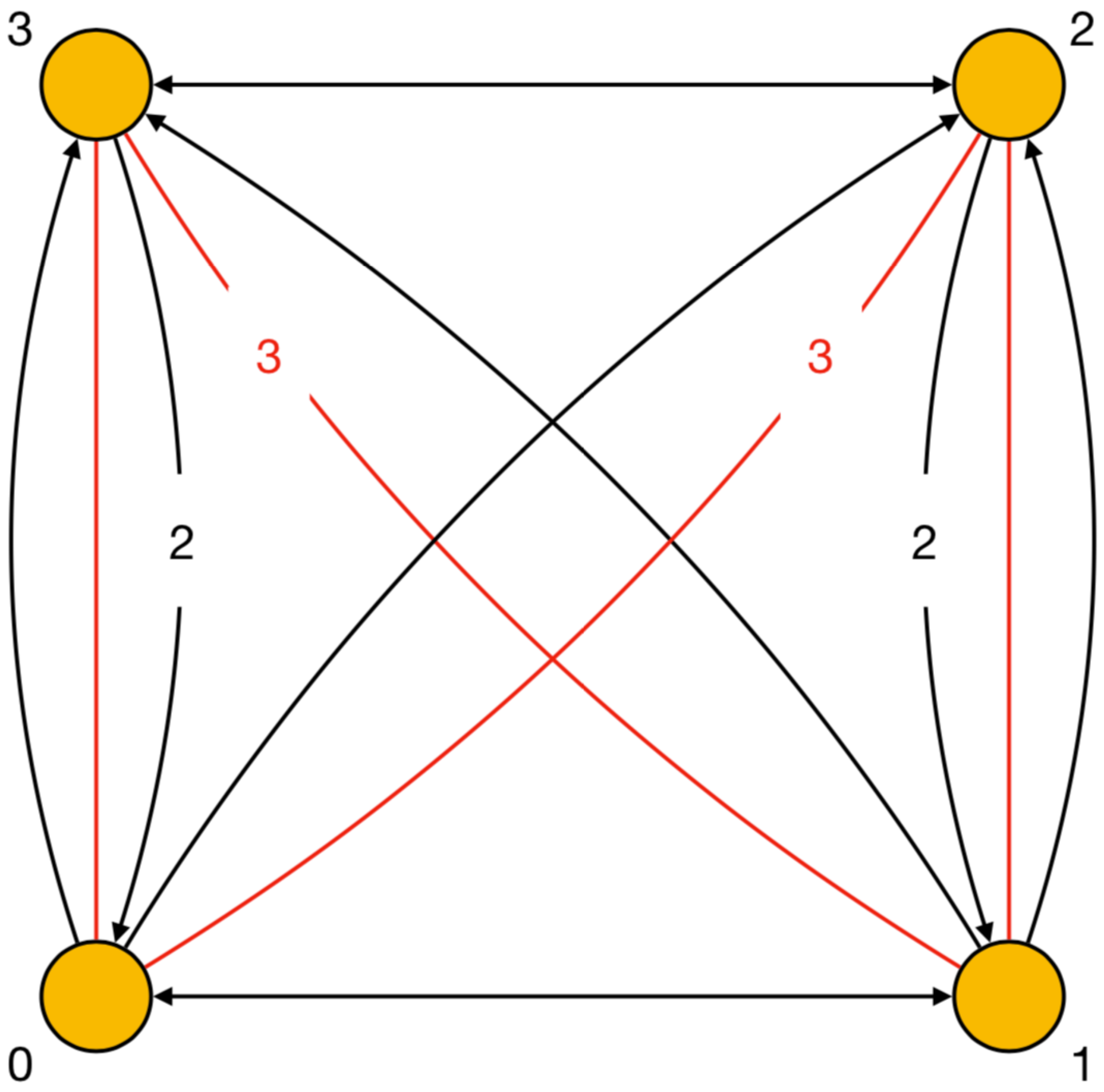}\quad
  \caption{Model 9}
  \label{fig:2}
\end{subfigure}
\medskip
\begin{subfigure}{0.4\textwidth}
  \centering
  \includegraphics[width=.7\linewidth]{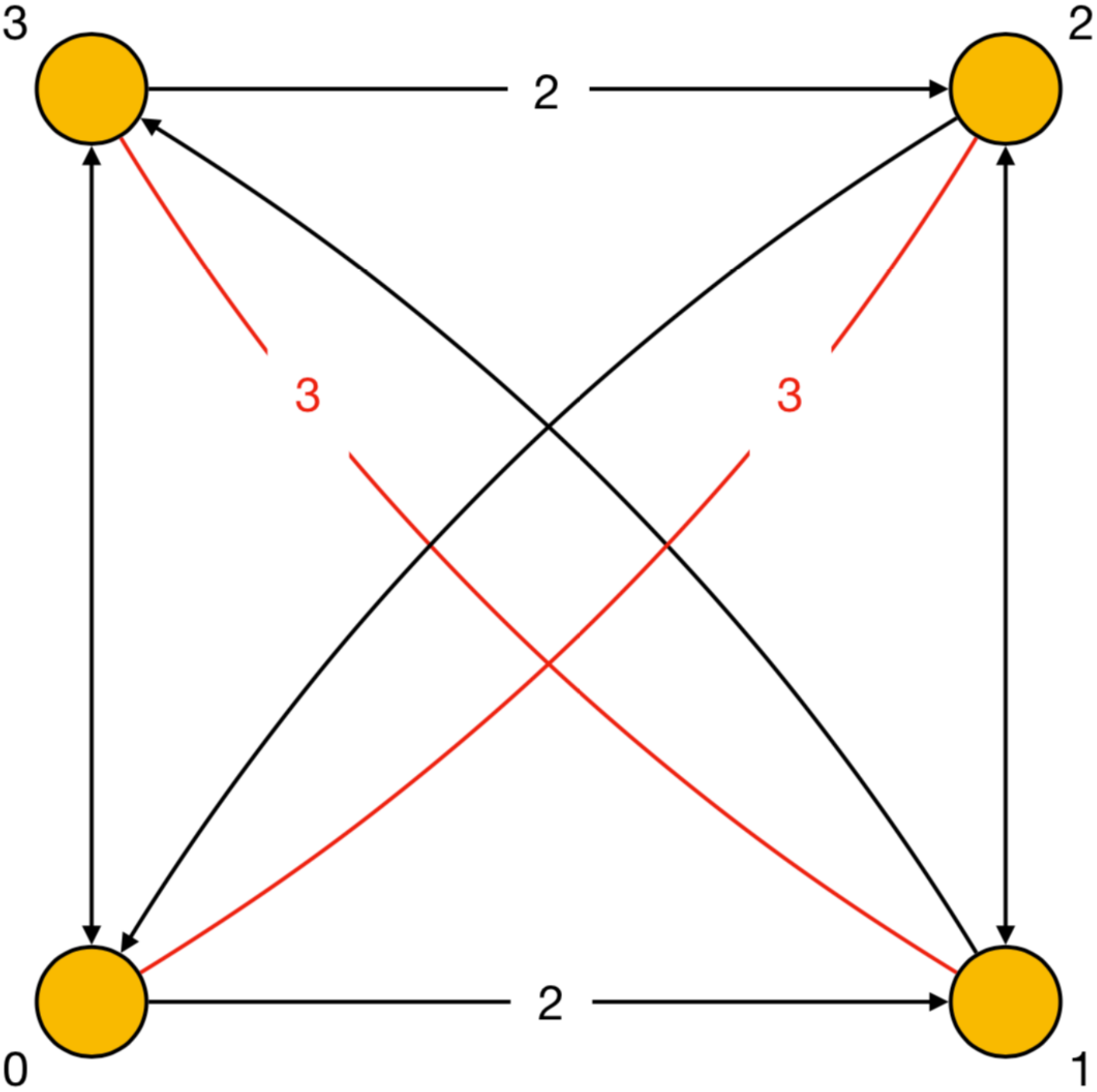}\quad
  \caption{Model 15, $H_4$}
  \label{fig:4}
\end{subfigure}
\begin{subfigure}{0.4\textwidth}
  \centering
  \includegraphics[width=.7\linewidth]{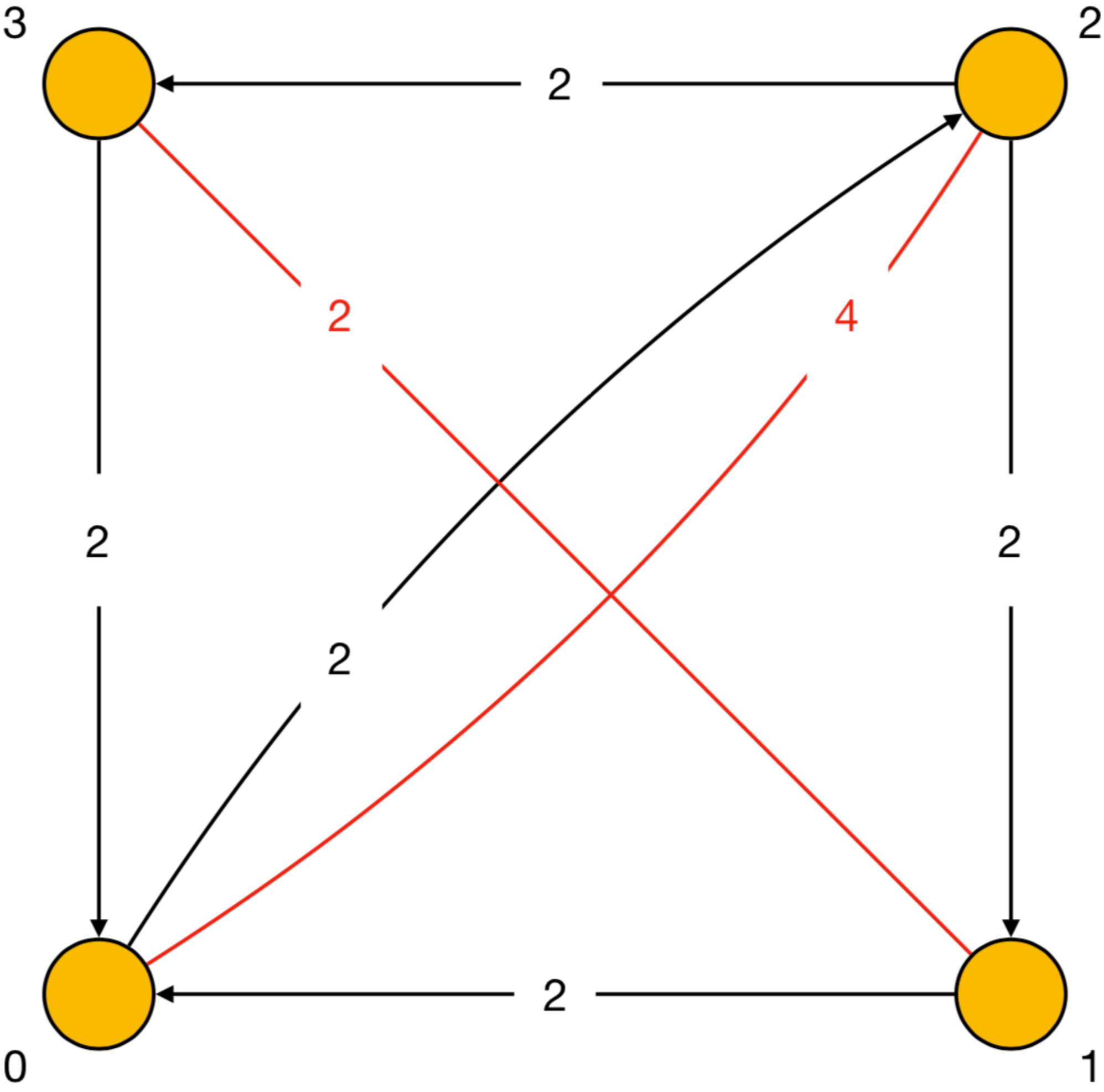}\quad
  \caption{Model 17, $Q^{1,1,1}$}
  \label{fig:5}
\end{subfigure}
\end{center}
\caption{Quivers associated to $S_2$ and with a toric CY$_4$ realization.}
\label{quivers_S2}
\end{figure}

%=================================================================
\paragraph{\underline{$S_2$}.}
%=================================================================
Interestingly, all the remaining toric theories that admit an interpretation as one of our seeds correspond to $S_2$. Their quivers are summarized in \fref{quivers_S2}, where we indicate the corresponding toric diagrams from \eref{toric_diagrams_volume_4}. The gauge theories for two of these geometries, usually referred to as $H_4$ and $Q^{1,1,1}$, have been previously studied in the literature \cite{Franco:2015tna,Franco:2016nwv,Franco:2017cjj,Franco:2018qsc}.

%=================================================================
%=================================================================

Since these quivers are not monochromatic, connecting to $S_2$ involves the cancellation of chiral-Fermi pairs, which contribute with opposite signs to \eref{S_matrix_quiver}. It is worth noting that all the quivers in \fref{quivers_S2} except for $Q^{1,1,1}$ involve chiral fields going in opposite directions between some pairs of nodes.

%=================================================================
\subsection{Duality Webs}
%=================================================================

\label{section_seeds_webs}

Duality webs map the space of dual theories and show how they are connected by mutations. In this section we present the duality webs for the seeds obtained in \sref{section_classification_seeds}, which are associated to the Diophantine equations for $\mathbb{C}^4/\mathbb{Z}_4$, \eref{diophantine_p4_z} and \eref{diophantine_p4_z2}. These are $m=2$ theories, namely $2d$ $\mathcal{N}=(0,2)$ gauge theories, so their webs involve triality.

Before focusing on these examples, let us discuss some general properties of duality webs for general $m$. Let us denote $N_{\mu}$ the number of gauge groups without adjoints in the quiver at a given site $\mu$ of the duality web. This number can differ between duals, namely it can change from site to site. At present it is unknown whether, in general, it is possible to dualize nodes with adjoint matter and, if so, how to do it. Therefore, for $m\geq 2$ the duality webs contain $N_{\mu}$ incoming arrows and $N_{\mu}$ outgoing arrows at site $\mu$. They correspond to acting with the mutation or the inverse mutation on every mutable node.\footnote{The distinction between mutation and inverse mutation is a matter of convention} For $m=1$, the mutation becomes the usual Seiberg duality, and for every node without adjoints in the quiver the mutation and its inverse collapse into a single, unoriented line.

In addition, every duality web contains length $(m-2)$ closed oriented loops, which correspond to dualizing $(m-2)$ consecutive times the same node of the quiver.\footnote{These loops are trivial for $m=1$. They correspond to going back and forth along an unoriented line in the web.} There are $N_{\mu}$ such loops passing through every site $\mu$ of the web. More interestingly, webs might contain closed loop associated to sequence of mutations involving more than one node of the quiver. Such loops are related to duality cascades.

Let us now return to the seeds under consideration. Remarkably, all these infinite theories give rise to just two distinct duality web structures. Of course, even though the structure of two webs associated to different seeds might coincide, they differ in the theories sitting at every site.

%=================================================================
\subsubsection*{Web 1: $S_t$ and $S_e$}
%=================================================================

\begin{figure}[ht]
	\centering
	\includegraphics[width=12cm]{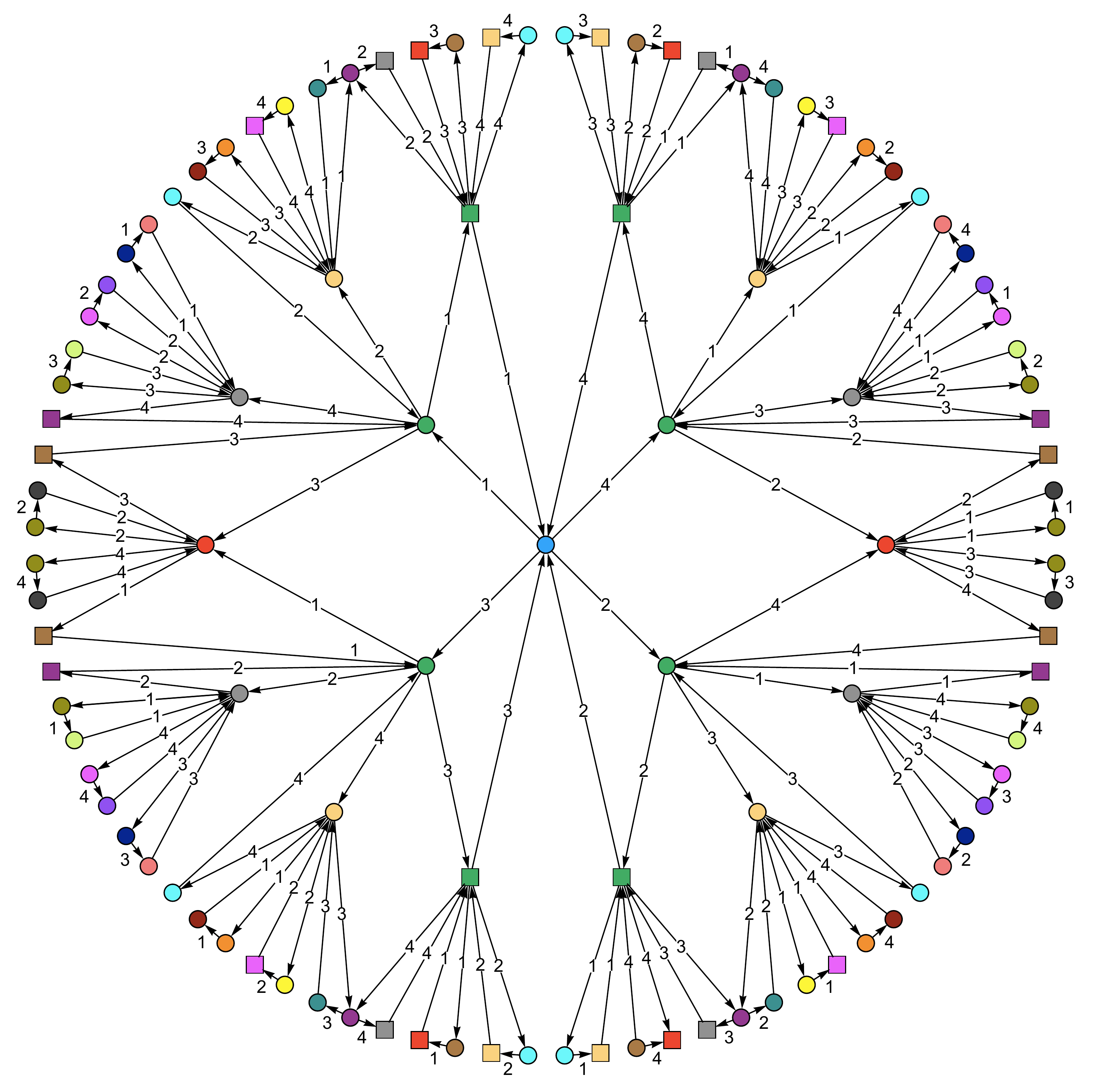}
\caption{Duality web for both $S_t$ and $S_e$. The web is infinite and we only present a portion of it here. The seed, i.e. either $S_t$ or $S_e$, corresponds to the blue circle at the center of the web. Over every arrow we indicate mutated node.}
	\label{duality_web_St}
\end{figure}

\fref{duality_web_St} shows the duality web for $S_t$, namely for $\mathbb{C}^4/\mathbb{Z}_4$. Different quivers, up to permutation of the nodes, are indicated with different shapes and colors. Distinct sites 
in the web with the same shape and color differ by a permutation of the nodes. The numbers on the arrows indicate the corresponding mutated node. In \eref{theories_on_St_duality_web} we present some of the $S$ matrices, after reordering of the nodes as explained in \sref{section_ordering_nodes}, encoding some of the quivers in the web, together with the ranks normalized by a factor $N$.
%=================================================================
\beq
\begin{array}{rclcrclcrcl}
S_{\includegraphics[width=1.5mm]{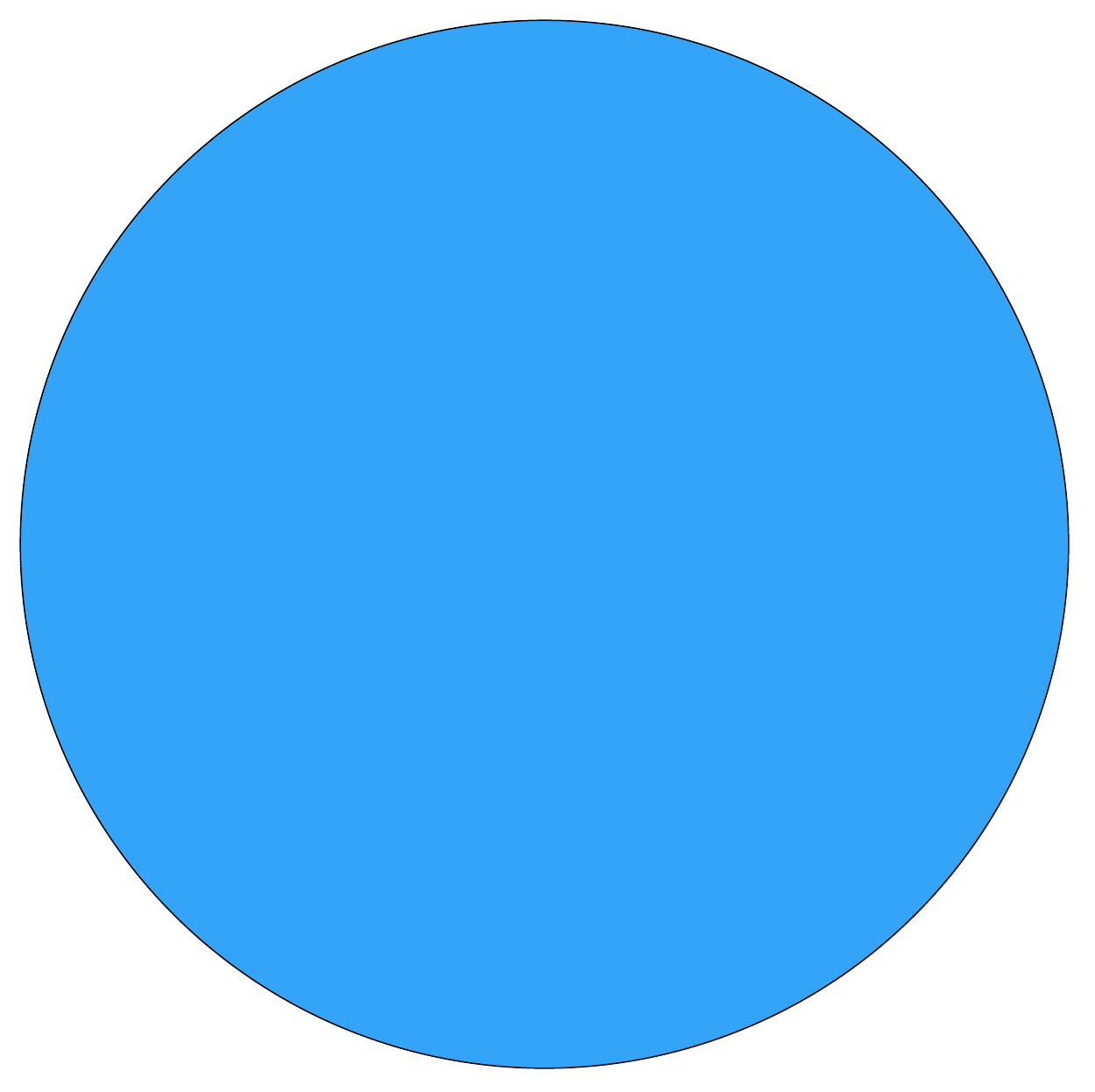}} & = & \begin{pmatrix}
                                1 & -4 & 6 & -4 \\
                                0 &  1 & -4 & 6 \\
                                0 & 0  & 1 & -4 \\
                                0 & 0 & 0 & 1
                            \end{pmatrix} & \ \ \ \ \ \ &
S_{\includegraphics[width=1.5mm]{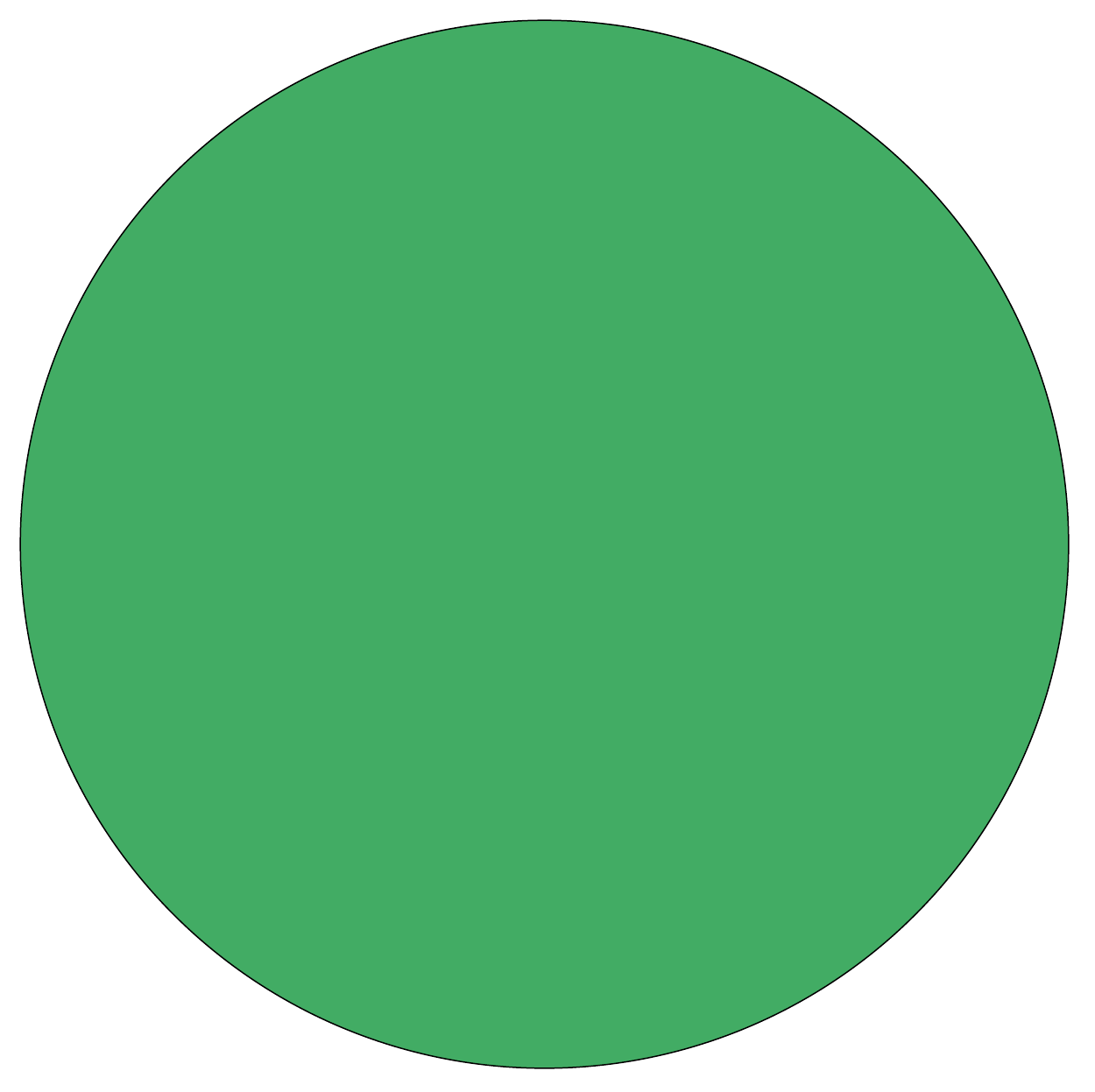}} & = & \begin{pmatrix}
                                1 & -4 & 4 & -10 \\
                                0 &  1 & -6 & 20 \\
                                0 & 0  & 1 & -4 \\
                                0 & 0 & 0 & 1
                            \end{pmatrix} 
                            & \ \ \ \ \ \ &
S_{\includegraphics[width=1.5mm]{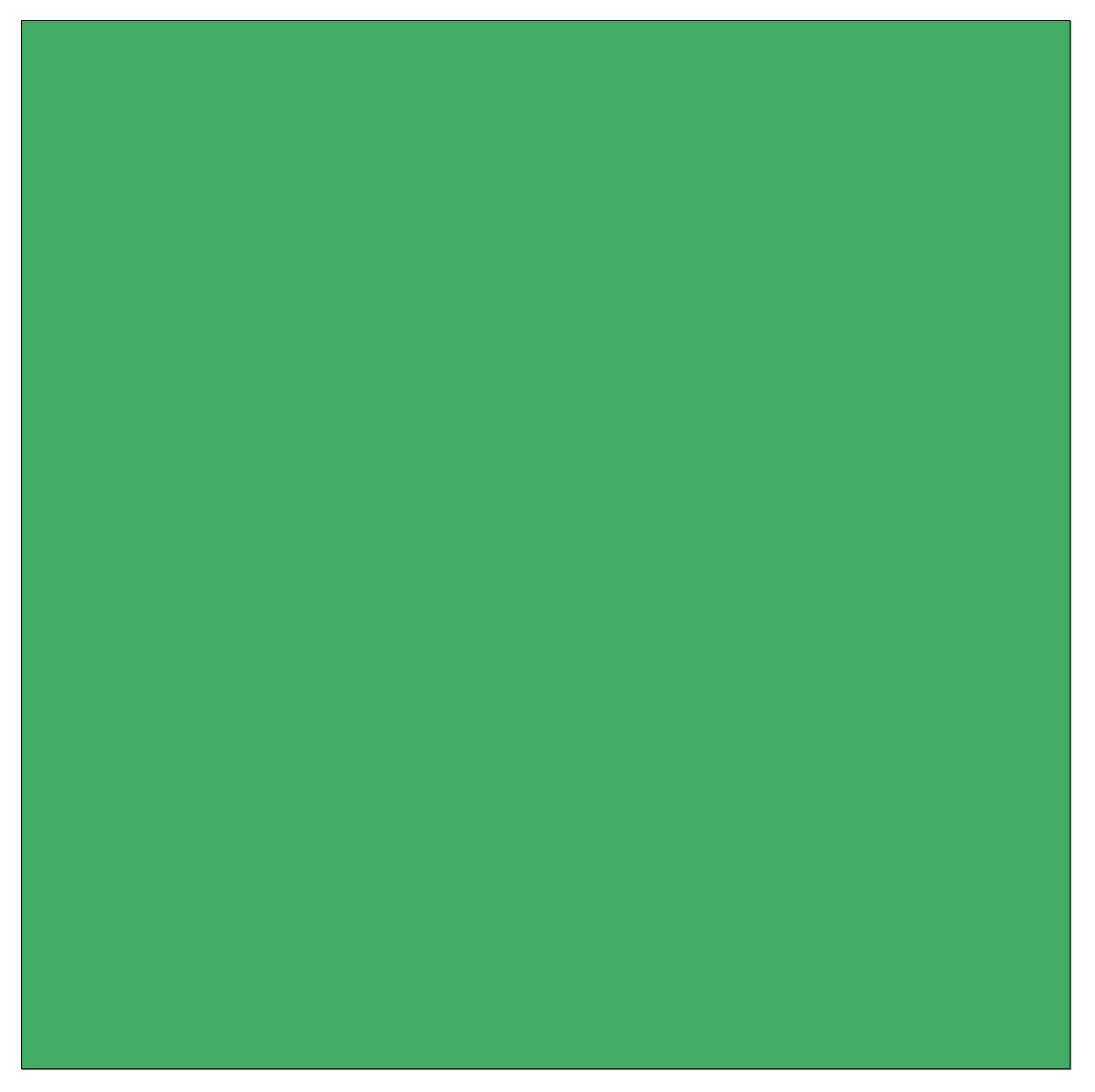}} & = & \begin{pmatrix}
                                1 & -4 & 20 & -10 \\
                                0 &  1 & -6 & 4 \\
                                0 & 0  & 1 & -4 \\
                                0 & 0 & 0 & 1
                            \end{pmatrix} \\ \\ 
\vec{N}_{\includegraphics[width=1.5mm]{web_quiver_0.pdf}} & = &(1,1,1,1) & \ \ \ \ \ \ &
\vec{N}_{\includegraphics[width=1.5mm]{web_quiver_1.pdf}} & = & (1,1,3,1) & \ \ \ \ \ \ &
\vec{N}_{\includegraphics[width=1.5mm]{web_quiver_2_1.pdf}} & = & (1,3,1,1) \\ \\ 
%%%%%%%%%%%%%%%%%%%%%%%%%%%%%%%
S_{\includegraphics[width=1.5mm]{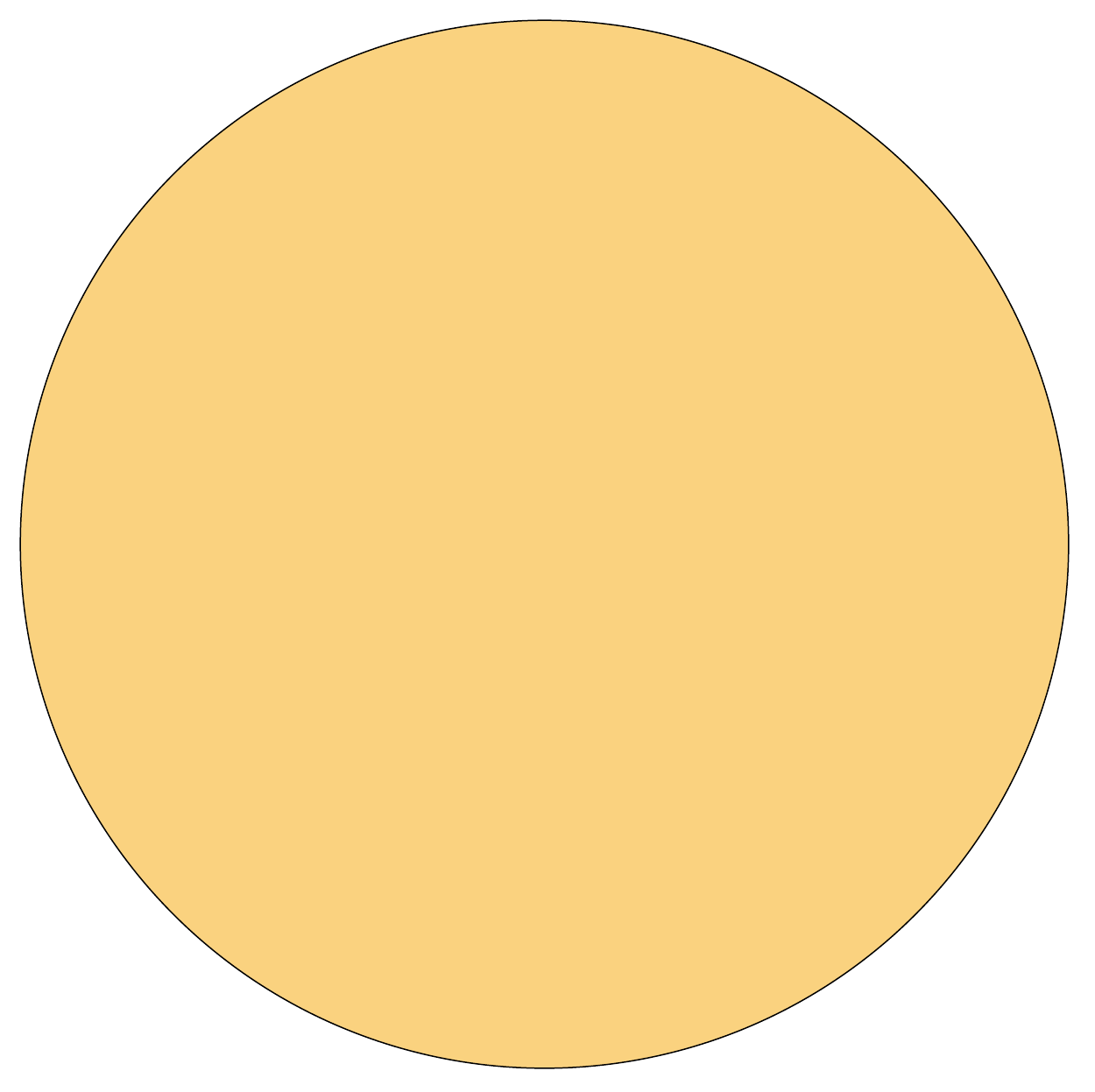}} & = & \begin{pmatrix}
                                1 & -10 & 4 & -4 \\
                                0 &  1 & -20 & 36 \\
                                0 & 0  & 1 & -6 \\
                                0 & 0 & 0 & 1
                            \end{pmatrix} & \ \ \ \ \ \ &
S_{\includegraphics[width=1.5mm]{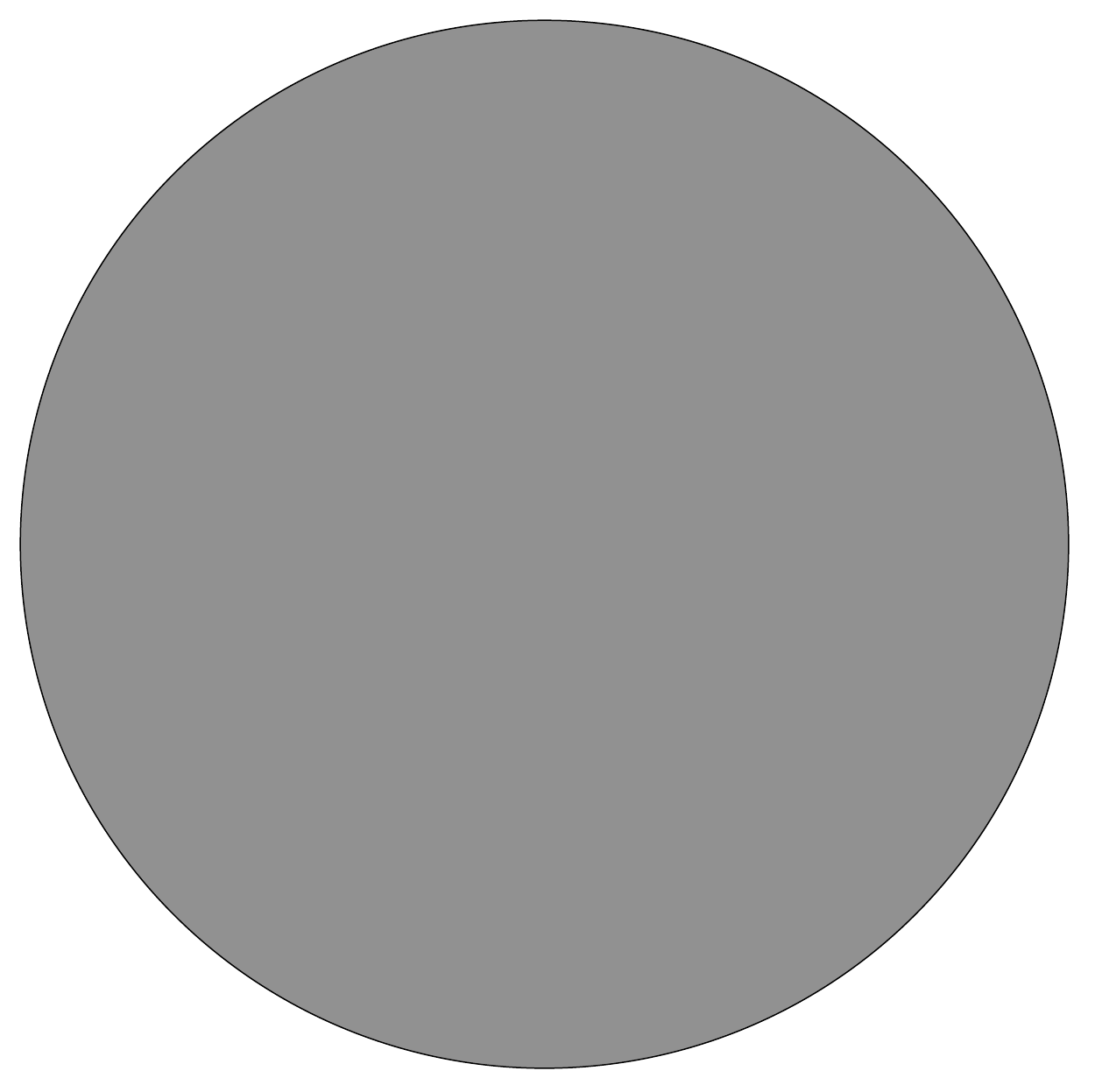}} & = & \begin{pmatrix}
                                1 & -4 & 10 & -36 \\
                                0 &  1 & -20 & 74 \\
                                0 & 0  & 1 & -4 \\
                                0 & 0 & 0 & 1
                            \end{pmatrix} 
                            & \ \ \ \ \ \ &
S_{\includegraphics[width=1.5mm]{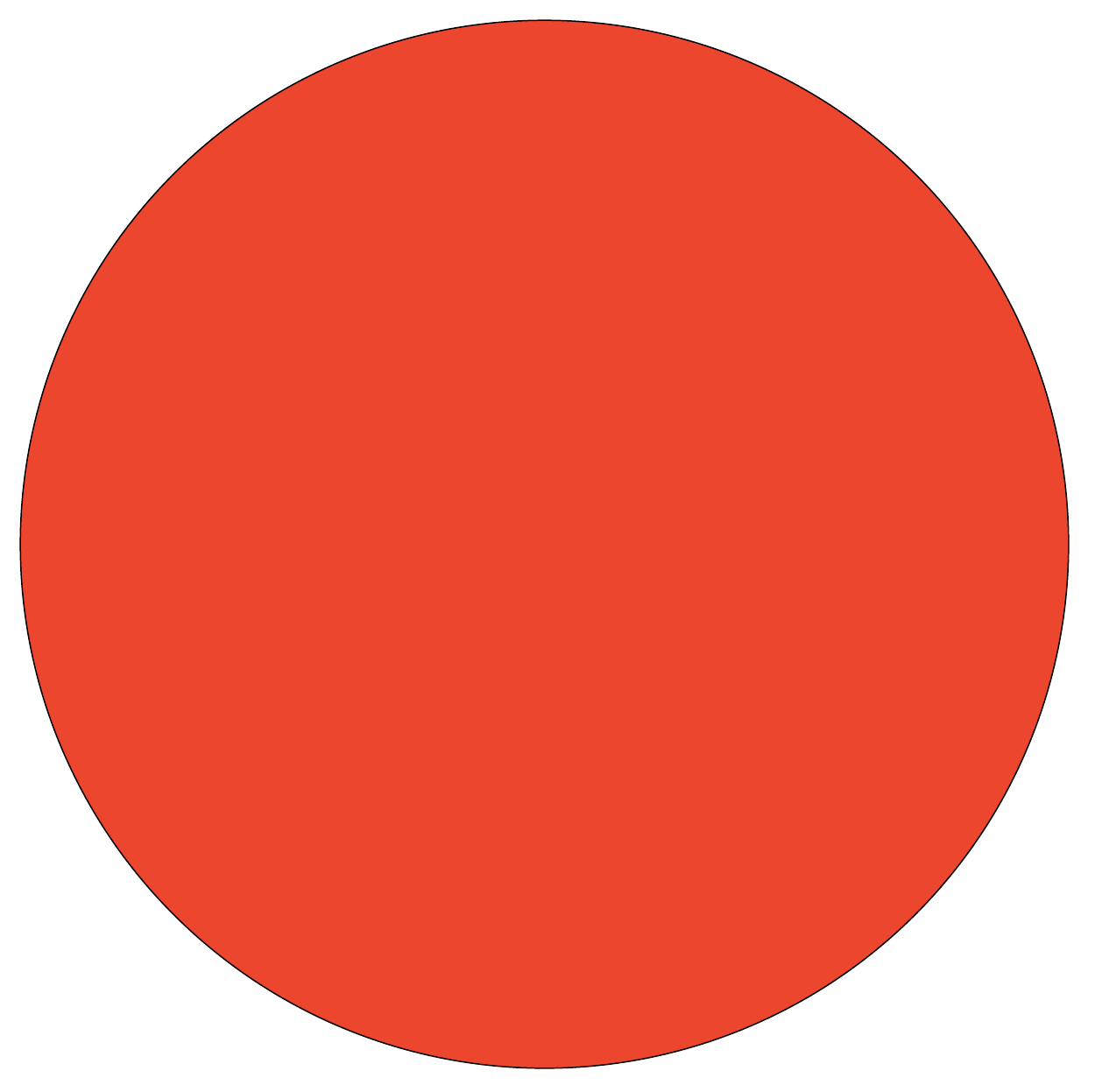}} & = & \begin{pmatrix}
                                1 & -20 & 70 & -4 \\
                                0 &  1 & -4 & 6 \\
                                0 & 0  & 1 & -20 \\
                                0 & 0 & 0 & 1
                            \end{pmatrix} \\ \\
\vec{N}_{\includegraphics[width=1.5mm]{web_quiver_2_2.pdf}} & = & (9,1,1,3) & \ \ \ \ \ \ &
\vec{N}_{\includegraphics[width=1.5mm]{web_quiver_2_3.pdf}} & = & (1,1,11,3) & \ \ \ \ \ \ &
\vec{N}_{\includegraphics[width=1.5mm]{web_quiver_2_4.pdf}} & = & (1,3,1,3)
\end{array}
\label{theories_on_St_duality_web}
\eeq
%=================================================================
In the web, we have chosen to distinguish theories that differ by conjugation of the chiral fields. This is a symmetry of $2d$ $(0,2)$ theories, since it is equivalent to conjugation of all bifundamental fields (chirals and Fermis) plus conjugation of the Fermis (which is a symmetry). The theories $\includegraphics[width=2mm]{web_quiver_1.pdf}$ and $\includegraphics[width=2mm]{web_quiver_2_1.pdf}$ are examples of this situation.

It is straightforward to verify that the theories in \eref{theories_on_St_duality_web} are anomaly free and satisfy the Diophantine equations \eref{diophantine_p4_z} and \eref{diophantine_p4_z2}.

%=================================================================
%=================================================================

Interestingly, the duality web for $S_e$ has the same structure of the one in \fref{duality_web_St}. The 4-fold symmetry of the web reflects the symmetry of the $S_t$ and $S_e$ seeds.

%=================================================================
\subsubsection*{Web 2: $S_n$}
%=================================================================

\begin{figure}[ht]
	\centering
	\includegraphics[width=12cm]{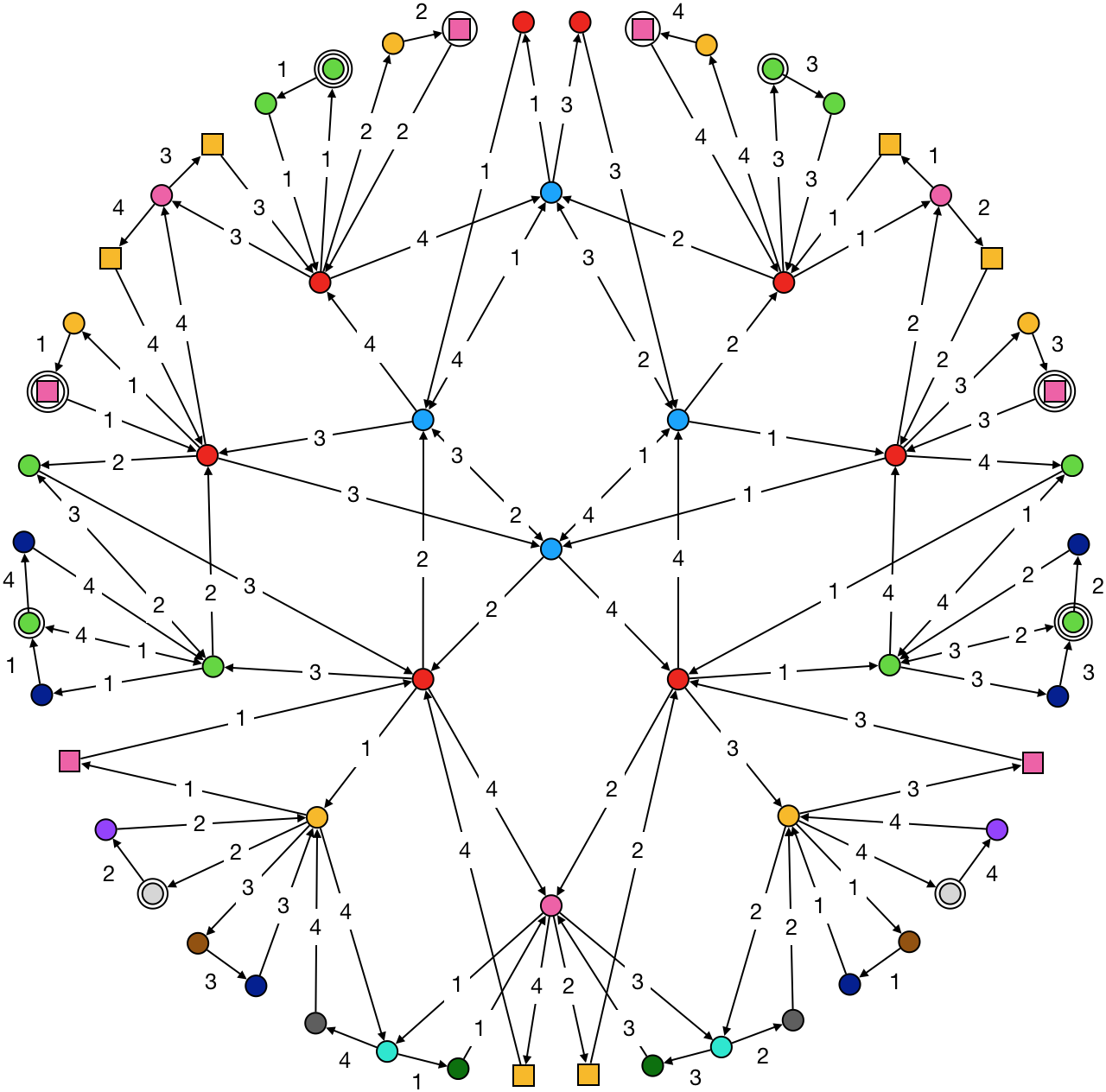}
\caption{Duality web for the $S_n$ seeds with $n>2$. The web is infinite and we only present a portion of it here. The seed, i.e. one of the $S_n$ theories, corresponds to the blue circle at the center of the web.}
	\label{duality_web_Sn}
\end{figure}

The duality webs for all the $S_n$ seeds with $n>2$ have the structure shown in \fref{duality_web_Sn}.\footnote{For $S_2$, the quiver is self-dual up to permutations of the nodes. The duality web is therefore rather trivial and simply consists of permutations of the seed.} The sites on the perimeter with the same symbol and number of surrounding circles are identified. 

This web exhibits a new feature: the presence of bidirectional arrows between some pairs of theories, for which going in opposite directions corresponds to triality on different nodes of the quiver. In this particular example, such bidirectional arrows only exist between quivers that differ just by a permutation of the nodes. It would be interesting to investigate whether this phenomenon is more general.

The reflection symmetry of the web with respect to the vertical axis follows from the symmetry of the $S_n$ seeds. For brevity, we do not present the $S$ matrices for different sites, but they can be easily constructed from the seed.

%=================================================================
%=================================================================

%=================================================================
\section{Periodic Cascades}
%=================================================================

\label{section_cascades}

As mentioned earlier, for any $m$, acting with $(m+1)$ consecutive mutations on a given node of a quiver takes us back to the original theory. Interestingly, more general periodic sequences of dualities can exist for arbitrary $m$. Explicit examples for $m>1$ have been presented in \cite{Franco:2016nwv,Closset:2018axq}. Borrowing the $m=1$ nomenclature, we refer to them as duality cascades.

\begin{figure}[ht]
	\centering
	\includegraphics[width=14cm]{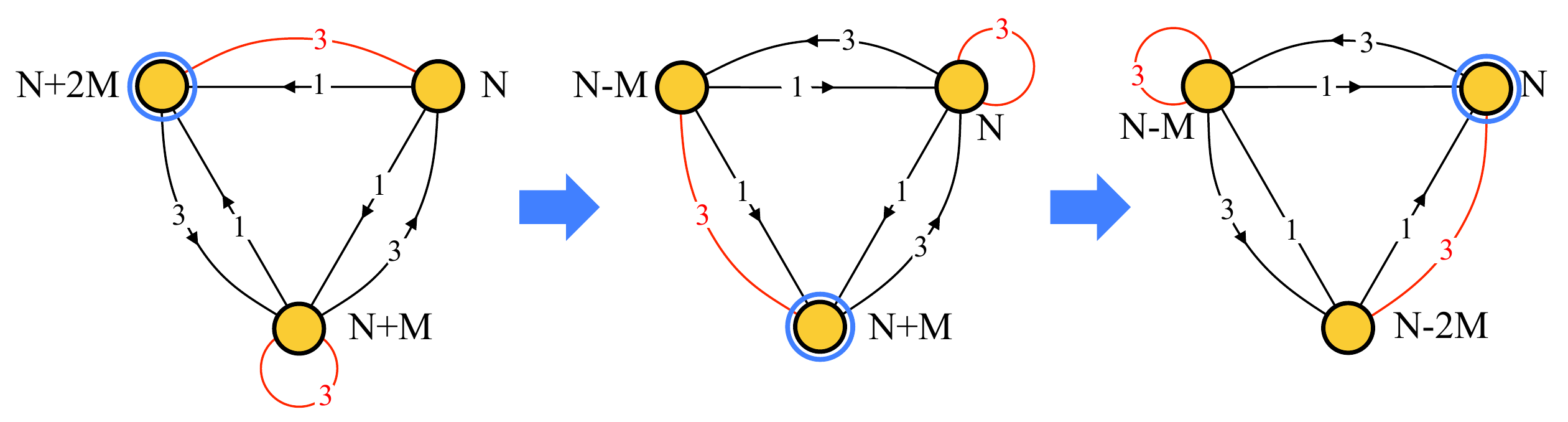}
\caption{Three stages in the duality cascade for $Y^{1,0}(\mathbb{P}^2)$ in the presence of regular and fractional branes. At every step we dualize the node with the highest rank, which is indicated in blue.}
	\label{cascade_Y10_P2}
\end{figure}

In the presence of regular and fractional branes, the number of regular branes often decreases with the dualizations. Moreover, in some cases, the number of fractional branes remains constant along this process. 
These two features indicate a reduction in the number of degrees of freedom and are analogous to the behavior of basic duality cascades for $m=1$.\footnote{More general behaviors are possible for $m=1$. In particular, the number of fractional branes can also change along cascades in theories with flavors (see e.g. \cite{Franco:2003ja,Ouyang:2003df,Franco:2003ea,Franco:2008jc}).} \fref{cascade_Y10_P2} shows an example of an $m=2$ duality cascade with this property, which is associated to the $Y^{1,0}(\mathbb{P}^2)$ geometry. The $Y^{1,0}(\mathbb{P}^m)$ is an infinite family of CY $(m+2)$-folds generalizing the conifold, which corresponds to the $m=1$ case. For every $m$ there is a single toric phase. This infinite family of quiver theories was introduced in \cite{Closset:2018axq}, to which we refer the reader for further details. For these geometries the cascade simply acts by rotating the quiver, while reducing the number of regular branes as the number of fractional branes stays fixed, as shown in \fref{cascade_Y10_P2}

%=================================================================
%=================================================================

In $4d$, duality cascades can be interpreted as a novel type of RG flow in which, as a function of the RG scale, gauge groups are Seiberg dualized every time they go to infinite gauge coupling \cite{Klebanov:2000hb,Strassler:2005qs}. The running of the gauge couplings is logarithmic and controlled by the NSVZ beta function \cite{Novikov:1983uc}. When the number of fractional branes is much smaller than the number of regular branes, they can be viewed as a small breaking of conformal invariance with respect to the CFT on the regular branes. The cascading RG flow interpretation is supported by a beautiful match with gravity duals, where it translates into warped throats \cite{Klebanov:2000hb,Herzog:2001xk,Herzog:2002ih,Franco:2004jz}. Such cascades and the dual throat geometries are a powerful ingredient for generating hierarchies in string theory.

We can therefore ask whether $2d$ $\mathcal{N}=(0,2)$ cascades also admit an RG interpretation (either for the type of theories considered in this paper or for others coming from different constructions). After these theories flow to infinite gauge coupling, the FI couplings for the gauge groups have a running controlled by a beta function that is similar to the NSVZ beta function \cite{Chen:2019eta}. A natural conjecture is that a $2d$ cascade is a sequence of triality transformations that act every time that an FI coupling becomes infinite. We plan to revisit this question in the future.

%=================================================================   
\section{Conclusions}
%=================================================================       

 \label{section_conclusions}
 
We investigated the $m$-graded quiver theories associated to CY $(m+2)$-folds and their order $(m+1)$ dualities. In particular, we studied how monodromies give rise to mutation invariants, which in turn can be formulated as Diophantine equations that characterize the space of dual theories associated to a given geometry. Our work considerably extends previous applications of these ideas, which were primarily focused on the $m=1$ case. Moreover, many of the earlier works considered a single equation per geometry, instead of the full set of independent equations arising from the expansion of the characteristic polynomial.

In order to illustrate these general ideas, we considered the explicit examples of $\mathbb{C}^{m+2}/\mathbb{Z}_{m+2}$ orbifolds. We performed a thorough analysis of $\mathbb{C}^{4}/\mathbb{Z}_{4}$, including a classification of the seeds for the corresponding pair of Diophantine equations. Interestingly, the number of seeds in this case is infinite. We further used this example to illustrate how powerful techniques developed in recent years make the identification of those extra seeds that have a realization in terms of D1-branes probing toric CY 4-folds possible. 

Finally, we commented on the possible generalization of duality cascades to arbitrary $m$.

%======================================================================  
\acknowledgments
%======================================================================  
 
We would like to thank C. Closset, G. Musiker and M. Porrati. for enjoyable discussions and related collaborations. The research of SF was supported by the U.S. National Science Foundation grants PHY-1820721 and DMS-1854179. AH was supported by INFN grant GSS (Gauge Theories, Strings and Supergravity). The three authors would like to thank the Simons Center for Geometry and Physics for hospitality during the completion of this work.

%======================================================================
\bibliographystyle{JHEP}
\bibliography{mybib}
%======================================================================

\end{document}